\documentclass[journal=jacsat,manuscript=article]{achemso}
\usepackage{chemformula} 
\usepackage[T1]{fontenc} 
\usepackage{comment}
\usepackage{algorithm}
\usepackage[noend]{algpseudocode}
\usepackage{tikz}
\usetikzlibrary{shapes, arrows}
\usepackage{caption}
\usepackage{graphicx}
\usepackage[font=small]{caption}
\usepackage[labelformat=empty, position=top]{subcaption}
\usepackage{amsfonts}

\author{Mohammad Rahbar}
\altaffiliation{Current affiliation: Technical University of Munich; TUM School of Natural Sciences, Department of Chemistry, Lichtenbergstr. 4, D-85748 Garching, Germany}
\author{Christopher J. Stein}
\email{christopher.stein@tum.de}
\affiliation{Theoretische Physik and CENIDE, Universität Duisburg-Essen, D-47048 Duisburg, Germany}
\altaffiliation{Current affiliation: Technical University of Munich; TUM School of Natural Sciences, Department of Chemistry, Lichtenbergstr. 4, D-85748 Garching, Germany}
\title
  {A statistical perspective on microsolvation }
\abbreviations{IR,NMR,UV}
\begin{document}

\begin{tocentry}

\includegraphics[width=1\textwidth]{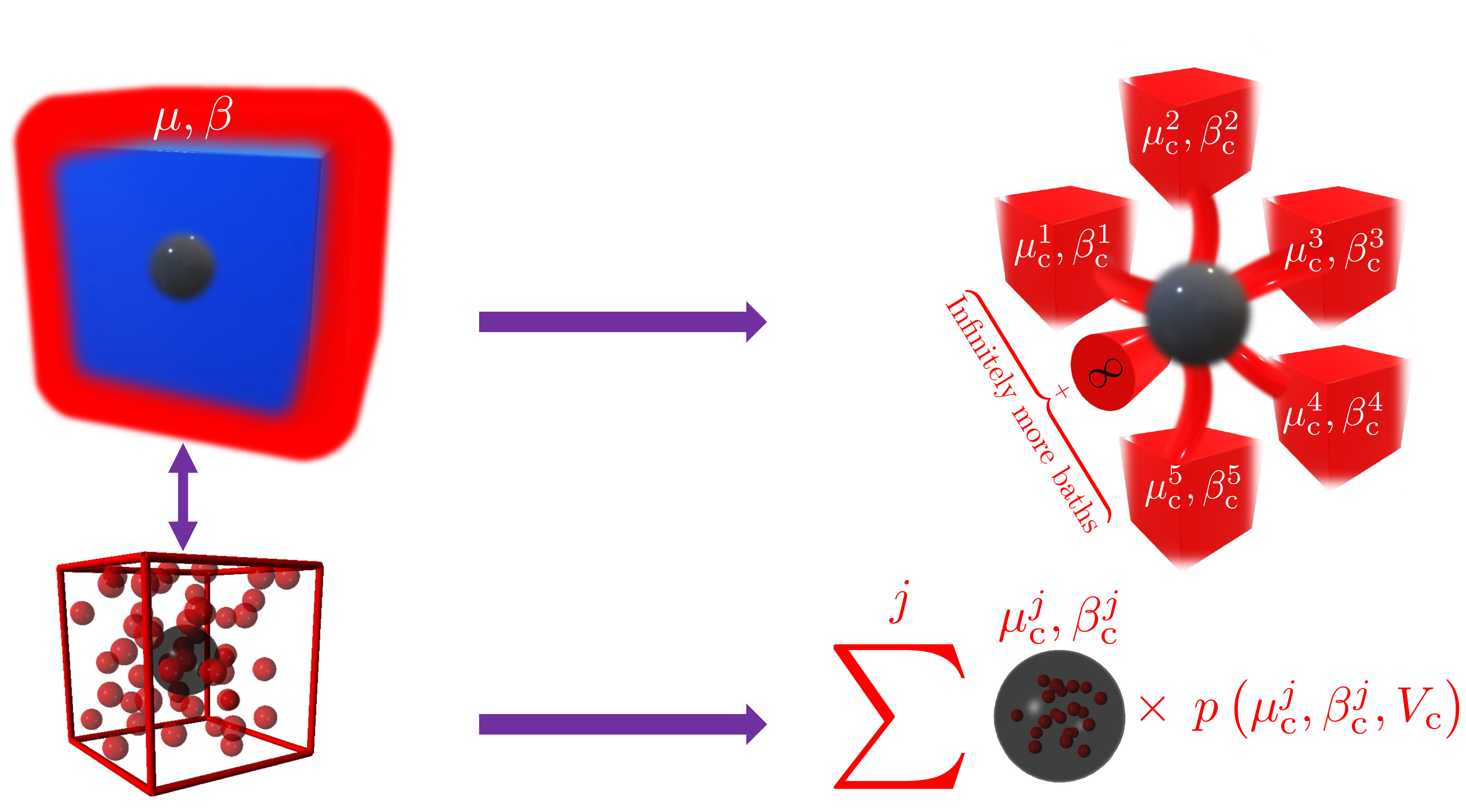}

\end{tocentry}

\begin{abstract}
The lack of a procedure to determine equilibrium thermodynamic properties of a small system interacting with a bath is frequently seen as a weakness of conventional statistical mechanics. A typical example for such a small system is a solute surrounded by an explicit solvation shell. 
One way to approach this problem is to enclose the small system of interest in a large bath of explicit solvent molecules, considerably larger than the system itself. The explicit inclusion of the solvent degrees of freedom is obviously limited by the available computational resources. A potential remedy to this problem is a microsolvation approach where only a few explicit solvent molecules are considered and surrounded by an implicit solvent bath.
Still, the sampling of the solvent degrees of freedom is challenging with conventional grand canonical Monte Carlo methods, since no single chemical potential for the solvent molecules can be defined in the realm of small-system thermodynamics.
 
In this work, a statistical thermodynamic model based on the grand canonical ensemble is proposed that avoids the conventional system size limitations and accurately characterizes the properties of the system of interest subject to the thermodynamic constraints of the bath. We extend an existing microsolvation approach to a generalized multi-bath "micro-statistical" model and show that the previously derived approaches result as a limit of our model. The framework described here is universal and we validate our method numerically for a Lennard--Jones model fluid.
\end{abstract}
\section{Introduction}
When we take the limit of $N \longrightarrow \infty$ particles, thermodynamics emerges from statistical mechanics. We can provide a microscopic justification for thermodynamics when we evaluate the relative fluctuations in thermodynamic quantities such as energy and show that they vanish as $1/ \sqrt N$.\cite{sethna2021statistical} When fluctuations are ignored, thermodynamics is the statistical mechanics of near-equilibrium systems.
However, this interpretation is no longer valid when a small system is examined. In other words, the lack of a procedure to determine equilibrium thermodynamic variables of a small system interacting with a bath is frequently seen as a weakness of conventional statistical mechanics.\cite{mandelbrot1989temperature,dunkel2014consistent,mandelbrot1962role,dixit2015detecting}
One way to deal with this problem is to directly simulate a small system enclosed in a bath in the thermodynamic limit, which typically requires the explicit inclusion of a large portion of the bath compared to the size of the small system. Such an approach is often constrained in practice by the available computational resources \cite{pratt1999quasi}. This is exacerbated by the generally high computational cost of accurate \textit{ab initio} calculations. For purely short-ranged interactions an often used conventional technique is to adopt periodic boundary conditions and empirically assess the system size dependence of the computed thermodynamical properties by carrying out simulations with successively larger baths. Typical problems in computational chemistry of small systems interacting with a large environment, such as molecule solvation or the study of nanoconfined species, however, can be substantially affected by long-range electrostatic interactions\cite{juffer1993dynamic,remsing2016,hegemann2017}. As a result, implicit solvent models that ignore the explicit molecular structure of the solvent bath in favor of an approximate description of its effect became increasingly popular over the last three decades.\cite{belch1985molecular,klamt1989,sato2000,mennucci2010,stein2019,herbert2021,ringe2022}   
Among the drawbacks of such implicit models is the fact that the accuracy of implicit solvent models relies strongly on the geometric details of the solute-solvent interface, which is often described by simple models rather than physical properties of the solute-solvent interactions \cite{thomas2013parameterization,lange2020,rana2022}.
Unlike implicit models, explicit models offer a physical and spatially
precise description of the solvent. However, as stated above, due to the sampling problem, many of the explicit models are computationally expensive and therefore may fail to reproduce experimental results. A potential remedy are hybrid methods that combine implicit and explicit schemes to reduce the computational cost while keeping the solvent’s spatial resolution in the solvation sell. Despite the hybrid approaches’ apparent effectiveness, there are unanswered questions
concerning the construction of microsolvation models\cite{li2013,kildgaard2018,simm2020,steiner2021} such as the subtleties of the coupling to the bulk solvent, and the extension of these models to solvent mixtures.

Understanding solvation phenomena hence requires quantitative knowledge of occupancy statistics of solvent molecules around the solutes. Recently, Dixit \textit{et al.} \cite{dixit2017mini} have shown that a single-bath grand canonical simulation does not allow for an accurate description when the size of the solvation shell is comparable to the size of the particle for a simple hard-sphere model system. 
In this article, we propose and derive a statistical thermodynamic model based on the grand canonical ensemble that avoids the conventional system size limitations of typical simulations with periodic boundary conditions by eliminating the need to explicitly include large parts of the thermodynamical bath in the simulations.
We extend an existing microsolvation approach to a generalized multi-bath "micro-statistical" model to obtain the properties of small systems and show that the previously derived approaches result as a limit of our model. We then validate our method numerically for a Lennard--Jones model fluid.
From a statistical point of view, we propose a model that allows us to sample the properties of a small \textit{core} system interacting with its bulk environment, the \textit{bath}, when the \textit{core} size is on the order of the average inter-particle  distance. This is achieved from a simulation of the \textit{core} alone.

\section{Theory}
\begin{figure}
    \centering
    \includegraphics[width=0.8\textwidth]{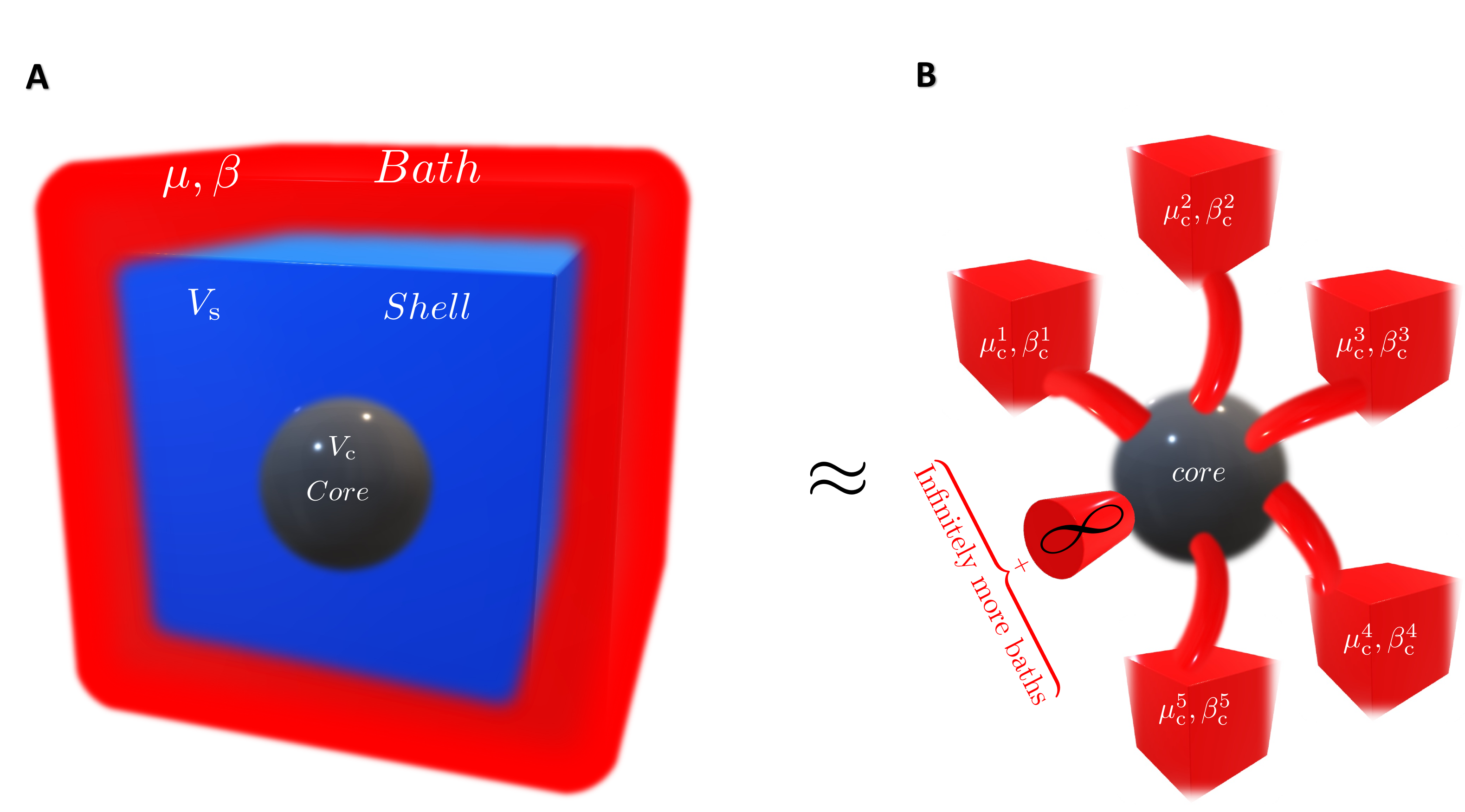}
    \caption{A) The composite system \textit{core}+\textit{shell} with the volume $V=V_{\text{c}}+V_{\text{s}}$, in which the volumes of the \textit{core} and \textit{shell} subsystems are $V_{\text{c}}$ and $V_{\text{s}}$, respectively, is in contact with a bath, at the chemical potential $\mu$ and the reciprocal
thermodynamic temperature  $\beta=1/k_\text{B} T$, where $k_\text{B}$ is the Boltzmann constant. $V_{\text{s}}$ is very large compared to $V_{\text{c}}$, and throughout this article, we consider $V	\simeq V_{\text{s}}$ and  $V>>V_{\text{c}}$. B) A schematic representation of mapping the problem of finding the microstate probability distribution for a small system to a system coupled to  theoretically (practically)  infinitely (finitely) many artificial thermodynamic
baths with the corresponding artificial thermodynamic variables, $\mu_\text{c}^j$ and $\beta_\text{c}^j$  (see Eq.~(\ref{eqpr})).}
    \label{fig2}
\end{figure}

Let us consider a system like the one shown on the left panel of Figure~\ref{fig2}. 
In the grand canonical ensemble, the composite system \textit{core+shell} is in contact with a bath and therefore
at constant chemical potential $\mu$ and  constant temperature $T$. The microstates of the combined system are characterized by the set $\{\Vec{\mathbf{q}},\Vec{\mathbf{p}}\} $, in which $\Vec{\mathbf{q}}$ and $\Vec{\mathbf{p}}$ are position and momenta of all particles, respectively. 
The probability distribution of the various microstates of this composite system \textit{core+shell} can be written as \cite{dixit2013maximum}
\begin{align}  p\left(\{\Vec{\mathbf{q}},\Vec{\mathbf{p}}\}_{V_{\mathrm{s}}}, \{\Vec{\mathbf{q}},\Vec{\mathbf{p}}\}_{V_{\mathrm{c}}} \mid \mu,\beta,V\right) 
\quad,\end{align}
with the volumes of the \textit{core} and \textit{shell} subsystem, $V_{\text{c}}$ and $V_{\text{s}}$, respectively and the reciprocal thermodynamic temperature $\beta=1/k_\text{B} T$ where $k_\text{B}$ is the Boltzmann constant.
The quantity $ p(\{\Vec{\mathbf{q}},\Vec{\mathbf{p}}\}_{V_{\mathrm{s}}}, \{\Vec{\mathbf{q}},\Vec{\mathbf{p}}\}_{V_{\mathrm{c}}} \mid \mu,\beta,V) $ can be interpreted as the joint probability distribution of microscopic degrees of freedom constrained by the macroscopic thermodynamic variables.  
So we can write,
\begin{align} 
&p\left(\{\Vec{\mathbf{q}},\Vec{\mathbf{p}}\}_{V_{\mathrm{s}}}, \{\Vec{\mathbf{q}},\Vec{\mathbf{p}}\}_{V_{\mathrm{c}}} \mid \mu,\beta,V\right)  
\nonumber\\&=\frac{\exp \left(\beta \mu N(\{\Vec{\mathbf{q}},\Vec{\mathbf{p}}\}_{V_{\mathrm{s}}}, \{\Vec{\mathbf{q}},\Vec{\mathbf{p}}\}_{V_{\mathrm{c}}})-\beta \mathcal{H}\left(\{\Vec{\mathbf{q}},\Vec{\mathbf{p}}\}_{V_{\mathrm{s}}}, \{\Vec{\mathbf{q}},\Vec{\mathbf{p}}\}_{V_{\mathrm{c}}}\right)\right)}{\mathcal{Q}\left(\mu,\beta,V\right)}   
\end{align}
with the grand partition function $\mathcal{Q}$, the total number of particles $N$ and the Hamiltonian of combined system $\mathcal{H}$.
The grand partition function is defined as 
\begin{align}
  &{\mathcal{Q}\left(\mu,\beta,V\right)}=\nonumber\\&\sum_{\{\Vec{\mathbf{q}},\Vec{\mathbf{p}}\}_{V_{\mathrm{s}}}, \{\Vec{\mathbf{q}},\Vec{\mathbf{p}}\}_{V_{\mathrm{c}}}} {\exp\left(\beta \mu N\left(\{\Vec{\mathbf{q}},\Vec{\mathbf{p}}\}_{V_{\mathrm{s}}}, \{\Vec{\mathbf{q}},\Vec{\mathbf{p}}\}_{V_{\mathrm{c}}}\right)
-\beta \mathcal{H}\left(\{\Vec{\mathbf{q}},\Vec{\mathbf{p}}\}_{V_{\mathrm{s}}}, \{\Vec{\mathbf{q}},\Vec{\mathbf{p}}\}_{V_{\mathrm{c}}}\right)\right)}
.\end{align}
Here and in the remainder of the article we write an integration over microstates as an infinite sum.
The Hamiltonian of the combined \textit{core+shell} system can be partitioned as follows
\begin{align}
\mathcal{H}\left(\{\Vec{\mathbf{q}},\Vec{\mathbf{p}}\}_{V_{\mathrm{s}}}, \{\Vec{\mathbf{q}},\Vec{\mathbf{p}}\}_{V_{\mathrm{c}}} \right) &\equiv \mathcal{H}_{\mathrm{s}}\left(\{\Vec{\mathbf{q}},\Vec{\mathbf{p}}\}_{V_{\mathrm{s}}}\right)+\mathcal{H}_{\mathrm{c}}\left(\{\Vec{\mathbf{q}},\Vec{\mathbf{p}}\}_{V_{\mathrm{c}}}\right) 
+\mathcal{H}_{{\mathrm{int}}}\left(\{\Vec{\mathbf{q}},\Vec{\mathbf{p}}\}_{V_{\mathrm{s}}}, \{\Vec{\mathbf{q}},\Vec{\mathbf{p}}\}_{V_{\mathrm{c}}}\right)
,\end{align}

where $\mathcal{H}_{\mathrm{c} (\mathrm{s})}$ is the Hamiltonian of the \textit{core(shell)} system, respectively and $\mathcal{H}_{\mathrm{int}}$ is the interaction between the subsystems. Our actual quantity of interest, the marginal distribution,\cite{dixit2013maximum} $ p( \{\Vec{\mathbf{q}},\Vec{\mathbf{p}}\}_{V_{\mathrm{c}}}\mid\mu,\beta,V)$, can be calculated by integrating the joint probability distribution, $p(\{\Vec{\mathbf{q}},\Vec{\mathbf{p}}\}_{V_{\mathrm{s}}},\{\Vec{\mathbf{q}},\Vec{\mathbf{p}}\}_{V_{\mathrm{c}}}\mid \mu,\beta,V)$, over all \textit{shell} degrees of freedom $\{\Vec{\mathbf{q}},\Vec{\mathbf{p}}\}_{V_{\mathrm{s}}}$

\begin{align}\label{eq1} 
& p\left( \{\Vec{\mathbf{q}},\Vec{\mathbf{p}}\}_{V_{\mathrm{c}}}\mid\mu,\beta,V\right)
=\sum_{\{\Vec{\mathbf{q}},\Vec{\mathbf{p}}\}_{V_{\mathrm{s}}}} p\left(\{\Vec{\mathbf{q}},\Vec{\mathbf{p}}\}_{V_{\mathrm{s}}}, \{\Vec{\mathbf{q}},\Vec{\mathbf{p}}\}_{V_{\mathrm{c}}}\mid ,\mu,\beta,V\right)\nonumber\\& \nonumber
 =\sum_{\{\Vec{\mathbf{q}},\Vec{\mathbf{p}}\}_{V_{\mathrm{s}}}} \frac{\exp \left(\beta \mu N(\{\Vec{\mathbf{q}},\Vec{\mathbf{p}}\}_{V_{\mathrm{s}}}, \{\Vec{\mathbf{q}},\Vec{\mathbf{p}}\}_{V_{\mathrm{c}}})-\beta \mathcal{H}\left(\{\Vec{\mathbf{q}},\Vec{\mathbf{p}}\}_{V_{\mathrm{s}}}, \{\Vec{\mathbf{q}},\Vec{\mathbf{p}}\}_{V_{\mathrm{c}}} \right)\right)}{\mathcal{Q}\left(\mu,\beta,V\right)} \\& \nonumber
={\exp \left(\beta\mu N( \{\Vec{\mathbf{q}},\Vec{\mathbf{p}}\}_{V_{\mathrm{c}}})-\beta \mu \mathcal{H}_{\mathrm{c}}\left(\{\Vec{\mathbf{q}},\Vec{\mathbf{p}}\}_{V_{\mathrm{c}}}\right) \right)}\\&\nonumber
\times\underbrace{\sum_{\{\Vec{\mathbf{q}},\Vec{\mathbf{p}}\}_{V_{\mathrm{s}}}}\frac{\exp \left(\beta \mu N(\{\Vec{\mathbf{q}},\Vec{\mathbf{p}}\}_{V_{\mathrm{s}}})-\beta \mathcal{H}_{\mathrm{s}}\left(\{\Vec{\mathbf{q}},\Vec{\mathbf{p}}\}_{V_{\mathrm{s}}}\right) -\beta\mathcal{H}_{{\mathrm{int} }}\left(\{\Vec{\mathbf{q}},\Vec{\mathbf{p}}\}_{V_{\mathrm{s}}}, \{\Vec{\mathbf{q}},\Vec{\mathbf{p}}\}_{V_{\mathrm{c}}}\right) \right)}{{\mathcal{Q}\left(\mu,\beta,V\right)}}}_{\mathcal{W }\left(\{\Vec{\mathbf{q}},\Vec{\mathbf{p}}\}_{V_{\mathrm{c}}} \mid \mu,\beta,V\right)}\nonumber\\& 
= {\exp  \left(\beta\mu N( \{\Vec{\mathbf{q}},\Vec{\mathbf{p}}\}_{V_{\mathrm{c}}})-\beta  \mathcal{H}_{\mathrm{c}}\left(\{\Vec{\mathbf{q}},\Vec{\mathbf{p}}\}_{V_{\mathrm{c}}}\right) \right) \mathcal{W }\left(\{\Vec{\mathbf{q}},\Vec{\mathbf{p}}\}_{V_{\mathrm{c}}} \mid \mu,\beta,V\right)}\, .
\end{align}
In this expression for the marginal distribution, the interaction between \textit{core} and \textit{shell} is exclusively described by $\mathcal{W }(\{\Vec{\mathbf{q}},\Vec{\mathbf{p}}\}_{V_{\mathrm{c}}}\mid \mu,\beta,V)$.
This interaction term is highly dependent on the characteristics of the \textit{core+shell} interaction Hamiltonian, $\mathcal{H}_{\text{int}}$, which in general is difficult to approximate. As a result, while Eq.~(\ref{eq1}) is strictly correct, it has no practical significance. An efficient method to calculate the distribution of the microstates of the \textit{core} system conditional to the thermodynamic constraints of the bath --- which is not in direct contact with the \textit{core} part of the system --- without the need of an explicit calculation of the interaction Hamiltonian $\mathcal{H}_{\text{int}}$ is therefore highly desired.

The need to explicitly calculate the interaction is rooted in a microstate interpretation of the joint probability distribution. One way to circumvent this issue is to introduce $j$ macrostate degrees of freedom and corresponding artificial thermodynamic variables, $\mu_\text{c}^j$ and $\beta_\text{c}^j$, and link these infinite macrostate degrees of freedom to the \textit{core} system microstates. This link is provided by the law of total probability, which can also be stated for conditional probabilities
\begin{align}\label{eqsix}
p\left( \{\Vec{\mathbf{q}},\Vec{\mathbf{p}}\}_{V_{\mathrm{c}}}\mid \mu,\beta,V\right) =\sum_{j} p\left( \{\Vec{\mathbf{q}},\Vec{\mathbf{p}}\}_{V_{\mathrm{c}}}\mid \mu,\beta,V,\mu^{j}_{\mathrm{c}},\beta^{j}_{\mathrm{c}},V_{\mathrm{c}}\right)p\left(  \mu^{j}_{\mathrm{c}},\beta^{j}_{\mathrm{c}},V_{\mathrm{c}}\mid\mu,\beta,V\right)    \, .
\end{align}
Now, using the chain rule for conditional probability (see the Appendix), we have
\begin{align}\label{eqseven}
&p\left( \{\Vec{\mathbf{q}},\Vec{\mathbf{p}}\}_{V_{\mathrm{c}}}\mid \mu,\beta,V,\mu^{j}_{\mathrm{c}},\beta^{j}_{\mathrm{c}},V_{\mathrm{c}}\right) 
=
\frac{p\left( \{\Vec{\mathbf{q}},\Vec{\mathbf{p}}\}_{V_{\mathrm{c}}}, \mu,\beta,V,\mu^{j}_{\mathrm{c}},\beta^{j}_{\mathrm{c}},V_{\mathrm{c}}\right) }{p\left( \mu,\beta,V,\mu^{j}_{\mathrm{c}},\beta^{j}_{\mathrm{c}},V_{\mathrm{c}}\right) }
\nonumber\\&=\frac{p\left( \mu^{j}_{\mathrm{c}},\beta^{j}_{\mathrm{c}},V_{\mathrm{c}}\right)p\left( \{\Vec{\mathbf{q}},\Vec{\mathbf{p}}\}_{V_{\mathrm{c}}}\mid\mu^{j}_{\mathrm{c}},\beta^{j}_{\mathrm{c}},V_{\mathrm{c}}\right)p\left(  \mu,\beta,V\mid\mu^{j}_{\mathrm{c}},\beta^{j}_{\mathrm{c}},V_{\mathrm{c}},\{\Vec{\mathbf{q}},\Vec{\mathbf{p}}\}_{V_{\mathrm{c}}}\right)}{p\left( \mu^{j}_{\mathrm{c}},\beta^{j}_{\mathrm{c}},V_{\mathrm{c}}\mid\mu,\beta,V\right)p\left( \mu,\beta,V\right)}\, .
\end{align}
For a sufficiently large \textit{shell} compared to a small \textit{core}, we can conclude that the newly introduced artificial thermodynamic constraints on the \textit{core} do not affect the thermodynamic properties of the \textit{shell} (which is still in equilibrium with the bath).  \cite{dixit2017mini} As a result, the artificial thermodynamic variables of the \textit{core} system and the \textit{shell} are almost independent.
With this assumption we arrive at
\begin{align}
p\left(  \mu^{j}_{\mathrm{c}},\beta^{j}_{\mathrm{c}},V_{\mathrm{c}}\mid\mu,\beta,V\right) \approx p\left(  \mu^{j}_{\mathrm{c}},\beta^{j}_{\mathrm{c}},V_{\mathrm{c}}\right)
\end{align}
and
\begin{align}\label{eqnine} 
p\left(  \mu,\beta,V\mid\mu^{j}_{\mathrm{c}},\beta^{j}_{\mathrm{c}},V_{\mathrm{c}},\{\Vec{\mathbf{q}},\Vec{\mathbf{p}}\}_{V_{\mathrm{c}}}\right)\approx p\left(  \mu,\beta,V\right)\, .
\end{align}
Inserting the results of Eqns. (\ref{eqseven})-(\ref{eqnine}) into Eq.~(\ref{eqsix}), we find
\begin{align}
&p\left( \{\Vec{\mathbf{q}},\Vec{\mathbf{p}}\}_{V_{\mathrm{c}}}\mid \mu,\beta,V\right)
\nonumber\\&\approx\sum_{j}
\frac{p\left( \mu^{j}_{\mathrm{c}},\beta^{j}_{\mathrm{c}},V_{\mathrm{c}}\right)p\left( \{\Vec{\mathbf{q}},\Vec{\mathbf{p}}\}_{V_{\mathrm{c}}}\mid\mu^{j}_{\mathrm{c}},\beta^{j}_{\mathrm{c}},V_{\mathrm{c}}\right)p\left(\mu,\beta,V\right)}{p\left( \mu^{j}_{\mathrm{c}},\beta^{j}_{\mathrm{c}},V_{\mathrm{c}}\right)p\left(\mu,\beta,V\right)}p\left(  \mu^{j}_{\mathrm{c}},\beta^{j}_{\mathrm{c}},V_{\mathrm{c}}\right)\nonumber\\&
\approx\sum_{j}p\left( \{\Vec{\mathbf{q}},\Vec{\mathbf{p}}\}_{V_{\mathrm{c}}}\mid \mu^{j}_{\mathrm{c}},\beta^{j}_{\mathrm{c}},V_{\mathrm{c}}\right) p\left(  \mu^{j}_{\mathrm{c}},\beta^{j}_{\mathrm{c}},V_{\mathrm{c}}\right)\, .\label{eqpr} 
\end{align}
This constitutes an approximation of the \textit{core} microstate distribution under the thermodynamical constraints of a bath which is not in direct contact with this \textit{core}. 
The first term in the sum, $p\left( \{\Vec{\mathbf{q}},\Vec{\mathbf{p}}\}_{V_{\mathrm{c}}}\mid \mu^{j}_{\mathrm{c}},\beta^{j}_{\mathrm{c}},V_{\mathrm{c}}\right)$, is the probability for a given \textit{core} microstate under the constraint of the artificial thermodynamic variables assigned to the \textit{core}, which is the standard grand canonical probability distribution for a chosen set of $\mu^j_\text{c}, \beta^j_\text{c}$ and $V_\text{c}$
\begin{align}\label{eqcon}
p\left( \{\Vec{\mathbf{q}},\Vec{\mathbf{p}}\}_{V_{\mathrm{c}}}\mid \mu^{j}_{\mathrm{c}},\beta^{j}_{\mathrm{c}},V_{\mathrm{c}}\right)  
=\frac{\exp \left(\beta^{j}_{\mathrm{c}}\mu^{j}_{\mathrm{c}} N(\{\Vec{\mathbf{q}},\Vec{\mathbf{p}}\}_{V_{\mathrm{c}}})-\beta^{j}_{\mathrm{c}}\mathcal{H}\left(\{\Vec{\mathbf{q}},\Vec{\mathbf{p}}\}_{V_{\mathrm{c}}}\right)\right)}{\mathcal{Q}\left(\mu^{j}_{\mathrm{c}},\beta^{j}_{\mathrm{c}},V_\text{c}\right)} \, .
\end{align}
We can interpret Eq.~(\ref{eqpr}) as a weighted average of the probability distributions of infinitely many artificial thermodynamic baths with the individual weights given by $p\left(  \mu^{j}_{\mathrm{c}},\beta^{j}_{\mathrm{c}},V_{\mathrm{c}}\right)$, Figure~\ref{fig2} (right panel). In the limit of the \textit{core} volume approaching the \textit{shell} volume, $\lim{V_\text{c}}\rightarrow V_\text{s}$, we expect that we can describe the probability distribution by a single bath with conventional grand canonical distribution function.
This seemingly obvious statement reveals an important property of the probability distribution $p(  \mu^{j}_{\mathrm{c}},\beta^{j}_{\mathrm{c}},V_{\mathrm{c}})$.
Taking this limit on both sides of Eq.~(\ref{eqpr}) we find
\begin{align}\label{eqd}
    \underbrace{\lim_{V_{\mathrm{c}}\rightarrow V}p\left( \{\Vec{\mathbf{q}},\Vec{\mathbf{p}}\}_{V_{\mathrm{c}}}\mid \mu,\beta,V\right)}_{p\left( \{\Vec{\mathbf{q}},\Vec{\mathbf{p}}\}_{V}\mid \mu,\beta,V\right)}&=\sum_{j}\lim_{V_{\mathrm{c}}\rightarrow V}p\left( \{\Vec{\mathbf{q}},\Vec{\mathbf{p}}\}_{V_{\mathrm{c}}}\mid \mu^{j}_{\mathrm{c}},\beta^{j}_{\mathrm{c}},V_{\mathrm{c}}\right) \lim_{V_{{\mathrm{c}}}\rightarrow V}p\left(  \mu^{j}_{\mathrm{c}},\beta^{j}_{\mathrm{c}},V_{\mathrm{c}}\right)\\&=\iint_{\mu_{\mathrm{c}}\beta_{\mathrm{c}}} \text{d}\mu_{\mathrm{c}} \text{d}\beta_{\mathrm{c}} p\left( \{\Vec{\mathbf{q}},\Vec{\mathbf{p}}\}_V\mid \mu_{\mathrm{c}},\beta_{\mathrm{c}},V\right)p\left(  \mu_{\mathrm{c}},\beta_{\mathrm{c}},V\right)\, .
\end{align}
We see that $p(  \mu_{\mathrm{c}},\beta_{\mathrm{c}},V)$ filters the value of the integrand $p( \{\Vec{\mathbf{q}},\Vec{\mathbf{p}}\}_V\mid \mu_{\mathrm{c}},\beta_{\mathrm{c}},V)$, at the point of its occurrence $(\mu,\beta)$ --- a property fulfilled by the Dirac delta function
\begin{align}\label{7} 
p\left(  \mu_{\mathrm{c}},\beta_{\mathrm{c}},V\right)\approx\delta^{2}(\mu_{\mathrm{c}}-\mu,\beta_{\mathrm{c}}-\beta)\, .
\end{align}

At this point we mapped the problem of finding the microstate probability distribution of a \textit{core} system coupled indirectly to a thermodynamic bath via a \textit{shell} that is in equilibrium with this bath, to a \textit{core} system coupled to infinitely many artificial thermodynamic baths (see Figure~\ref{fig2}). We are left with the difficulty of calculating the coupling strengths (or weights), $p\left(  \mu^{j}_{\mathrm{c}},\beta^{j}_{\mathrm{c}},V_{\mathrm{c}}\right)$, of the individual baths. However, in light of the thermodynamic limit, we require the weight distribution to collapse to a Dirac delta function for large \textit{core} volumes. 
We arrived at this interpretation without any reference to characteristic properties of the system. The above-mentioned behaviour at the thermodynamic limit was predicted in previous works \cite{bansal2017thermodynamics,dixit2017mini,bansal2016structure} but has not been rigorously proven so far.

\subsection{Identifying a functional form for the weight distribution}
The function $p(\mu^{j}_\text{c},\beta^{j}_\text{c},V_\text{c})$ assigns a weight to a bath with thermodynamic variables $\mu^{j}_\text{c}$ and $\beta^{j}_\text{c}$ coupled to the \textit{core} of size $V_\text{c}$. Unfortunately, the functional form of the weight function is unknown \textit{a priori}.
We now attempt to rewrite this unknown function in a more familiar form. Therefore, we start by applying the fundamental law of probability, meaning that for a given set of artificial thermodynamic variables $(\mu^{j}_\text{c},\beta^{j}_\text{c},V_\text{c})$, the sum of all the probabilities $p\left( \{\Vec{\mathbf{q}},\Vec{\mathbf{p}}\}_{V_{\mathrm{c}}}\mid \mu^{j}_{\mathrm{c}},\beta^{j}_{\mathrm{c}},V_{\mathrm{c}}\right)$ on all possible degrees of freedom ${\{\Vec{\mathbf{q}},\Vec{\mathbf{p}}\}_{V_\text{c}}}$ must be equal to one,
\begin{align}\label{fun}
\sum_{\{\Vec{\mathbf{q}},\Vec{\mathbf{p}}\}_{V_{\mathrm{c}}}} p\left( \{\Vec{\mathbf{q}},\Vec{\mathbf{p}}\}_{V_{\mathrm{c}}}\mid \mu^{j}_{\mathrm{c}},\beta^{j}_{\mathrm{c}},V_{\mathrm{c}}\right)=1\, .
\end{align}
Multiplying by $p(  \mu^{j}_\text{c},\beta^{j}_\text{c},V_\text{c})$ on both sides of Eq.~(\ref{fun}),
\begin{align}
    p\left(  \mu^{j}_{\mathrm{c}},\beta^{j}_{\mathrm{c}},V_{\mathrm{c}}\right)=p\left(  \mu^{j}_{\mathrm{c}},\beta^{j}_{\mathrm{c}},V_{\mathrm{c}}\right)\sum_{\{\Vec{\mathbf{q}},\Vec{\mathbf{p}}\}_{V_{\mathrm{c}}}} p\left( \{\Vec{\mathbf{q}},\Vec{\mathbf{p}}\}_{V_{\mathrm{c}}}\mid \mu^{j}_{\mathrm{c}},\beta^{j}_{\mathrm{c}},V_{\mathrm{c}}\right)
\end{align}
and using Eq.~(\ref{eqcon}),
we can write
\begin{align}
&p\left(  \mu^{j}_{\mathrm{c}},\beta^{j}_{\mathrm{c}},V_{\mathrm{c}}\right)
\nonumber\\&={\mathcal{Q}^{-1}\left(\mu^{j}_{\mathrm{c}},\beta^{j}_{\mathrm{c}},V_{\mathrm{c}}\right)}\underbrace{p\left(\mu^{j}_{\mathrm{c}},\beta^{j}_{\mathrm{c}},V_{\mathrm{c}}\right)\sum_{\{\Vec{\mathbf{q}},\Vec{\mathbf{p}}\}_{V_{\mathrm{c}}}}{\exp  \left(\beta^{j}_{\mathrm{c}} \mu^{j}_{\mathrm{c}} N( \{\Vec{\mathbf{q}},\Vec{\mathbf{p}}\}_{V_{\mathrm{c}}})-\beta^{j}_{\mathrm{c}}  \mathcal{H}_{\mathrm{c}}\left(\{\Vec{\mathbf{q}},\Vec{\mathbf{p}}\}_{V_{\mathrm{c}}}\right) \right) }}_{X\left(\mu^{j}_{\mathrm{c}},\beta^{j}_{\mathrm{c}},V_{\mathrm{c}}\right)}
\end{align}
By now, we can interpret $p\left(  \mu^{j}_{\mathrm{c}},\beta^{j}_{\mathrm{c}},V_{\mathrm{c}}\right)$ as a weighted distribution with a known part (the inverse of the grand canonical partition function) and a yet to be determined part $X\left(\mu^{j}_{\mathrm{c}},\beta^{j}_{\mathrm{c}},V_{\mathrm{c}}\right)$
\begin{align}
    p\left(  \mu^{j}_{\mathrm{c}},\beta^{j}_{\mathrm{c}},V_{\mathrm{c}}\right)=\mathcal{Q}^{-1}\left(\mu^{j}_{\mathrm{c}},\beta^{j}_{\mathrm{c}},V_{\mathrm{c}}\right) X\left(\mu^{j}_{\mathrm{c}},\beta^{j}_{\mathrm{c}},V_{\mathrm{c}}\right)\, .
\end{align}
For notational simplicity, we replace $(\mu^{j}_{\mathrm{c}},\beta^{j}_{\mathrm{c}},V_{\mathrm{c}})$ with $y$
\begin{align}
p\left(y\right)=\mathcal{Q}^{-1}\left(y\right) X\left(y\right)\, .
\end{align}

We know that $\mathcal{Q}^{-1}(y)$ has an exponential form with known argument $g(y)$. We can write the argument as the sum of two parts $g_1(y)$ and $g_2(y)$, respectively. It is trivial that after choosing one part (for which we have infinitely many choices), the other part is uniquely determined. Now we can write 
\begin{align}
p\left(y\right)= \exp\left({g_1\left(y\right)}+g_2\left(y\right)\right)  X\left(y\right)=  \exp\left({g_1\left(y\right)}+g_2\left(y\right)+
    \ln{X\left(y\right)}\right)\, .
\end{align}
 The idea is that among all of the infinite number of options for picking $g_2(y)$, we choose $X(y)$ for our ansatz, i.e. $X(y)=g_2(y)$, leading to
\begin{align}\label{unkn}
p\left(y\right)=\exp\left({\overbrace{g_1\left(y\right)}^{\text{determined}}}+\overbrace{g_2\left(y\right)}^{\text{choose}}+\ln{\overbrace{g_2\left(y\right)}^{\text{ansatz}}}\right)\, .
\end{align}
As is evident from Eq.~(\ref{unkn}), $p(y)$ is  fully defined.   
With the original variables we find
\begin{align}\label{eqQ}
p\left(  \mu^{j}_{\mathrm{c}},\beta^{j}_{\mathrm{c}},V_{\mathrm{c}}\right)\nonumber&=\mathcal{Q}^{-1}\left(g_1\left(\mu^{j}_{\mathrm{c}},\beta^{j}_{\mathrm{c}},V_{\mathrm{c}}\right),g_2\left(\mu^{j}_{\mathrm{c}},\beta^{j}_{\mathrm{c}},V_{\mathrm{c}}\right)\right) g_2\left(\mu^{j}_{\mathrm{c}},\beta^{j}_{\mathrm{c}},V_{\mathrm{c}}\right)
\nonumber\\&={\mathcal{Q}^{-1}\left(g_1\left(\mu^{j}_{\mathrm{c}},\beta^{j}_{\mathrm{c}},V_{\mathrm{c}}\right),g_2\left(\mu^{j}_{\mathrm{c}},\beta^{j}_{\mathrm{c}},V_{\mathrm{c}}\right)+\ln{g_2\left(\mu^{j}_{\mathrm{c}},\beta^{j}_{\mathrm{c}},V_{\mathrm{c}}\right)}\right)}\, .
\end{align}
Now the proposed functional form is proportional to the inverse of the partition function, which has a physical interpretation.
In summary, the functional form of the grand partition function has been used as a guide for constructing the weighted distribution. The remaining question is which part of the inverse partition function should be chosen as $g_2(y)$ and hence define $X(y)$. To identify a potential expression, we first need to inspect the grand partition function more closely. This is the topic of the next section.

\subsection{Grand canonical partition function for an interacting system of particles} 

The grand canonical partition function ${\mathcal{Q}(\beta^{j}_{\mathrm{c}}, \mu^{j}_{\mathrm{c}}, V_{\mathrm{c}})}$ can be written as a sum over canonical partition functions $\mathcal{Z}(\beta^{j}_{\mathrm{c}}, N^{'}, V_{\mathrm{c}})$
\begin{align}\label{eqq} 
\mathcal{Q}(\beta^{j}_{\mathrm{c}}, \mu^{j}_{\mathrm{c}}, V_{\mathrm{c}})=\sum_{N^{'}=0}^{\infty} \mathrm{e}^{\beta^{j}_{\mathrm{c}} \mu^{j}_{\mathrm{c}} N^{'}} \sum_{\left(\{\Vec{\mathbf{q}},\Vec{\mathbf{p}}\}_{V_{\mathrm{c}}} \mid N^{'}, V_{\mathrm{c}}  \right)} \mathrm{e}^{-\beta^{j}_{\mathrm{c}} \mathcal{H}_{N^{'}}\left(\{\Vec{\mathbf{q}},\Vec{\mathbf{p}}\}_{V_{\mathrm{c}}}\right)}=\sum_{N^{'}=0}^{\infty} \mathrm{e}^{\beta^{j}_{\mathrm{c}} \mu^{j}_{\mathrm{c}} N^{'}}\mathcal{Z}(\beta^{j}_{\mathrm{c}}, N^{'}, V_{\mathrm{c}})\, .
\end{align}
It is obvious that if we evaluate the canonical partition function $\mathcal{Z}(\beta^{j}_{\mathrm{c}}, N^{'}, V_{\mathrm{c}})$, then finding the grand canonical partition functions is straightforward. As discussed previously, we attempt to understand the role of more complicated interactions between particles such as the Lennard--Jones model and investigate how to treat them in our proposed framework.
We consider a one-component system of $N^{'}$ spherical particles with mass $m,$ positions $\Vec{\mathbf{q}}$ and momenta $\Vec{\mathbf{p}}$. Particles $i$ and $j$ interact via a pair-wise potential $\phi_{ij}\left(\mathbf{r}_{i}, \mathbf{r}_{j}\right)$. The Hamiltonian of this system of interacting particles can be written as,
\begin{align}\label{9} 
\mathcal{H}_{N^{'}}=\sum_{i=1}^{N^{'}} \frac{\vec{{p}_{i}}^{2}}{2 m}+\phi\left(\vec{q}_{1}, \cdots, \vec{q}_{N^{'}}\right)\, ,
\end{align}
in which $\phi\left(\vec{q}_{1}, \cdots, \vec{q}_{N}\right)$ is total internal potential function which can be written as
\begin{align}\label{potential} 
\phi\left(\vec{q}_{1}, \cdots, \vec{q}_{N^{'}}\right)=\frac{1}{2} \sum_{i \neq j}^{N^{'}} \phi_{ij}\left({\vec{q}}_{i}, {\vec{q}}_{j}\right)\, .
\end{align}
We can now write the canonical partition function of the system of interacting particles \cite{kardar2007statistical}. (see the Appendix for details).
\begin{align}\label{eqz}
 \mathcal{Z}(\beta^{j}_{\mathrm{c}}, N^{'}, V_{\mathrm{c}})= {\frac{V_{\mathrm{c}}^{N^{'}}}{N^{'} !}\left[\frac{\sqrt{2 m k_{B} T^{j}_{\mathrm{c}}}}{h}\right]^{3N^{'}}}{\left\langle\exp \left[-\beta^{j}_{\mathrm{c}} \phi\left(\vec{q}_{1}, \cdots, \vec{q}_{N^{'}}\right)\right]\right\rangle_{\textbf{Non-Int}}}    
\end{align}
Taking the natural logarithm on both sides of Eq.~(\ref{eqz}),
\begin{align}
\ln\mathcal{Z}(\beta^{j}_{\mathrm{c}}, N^{'}, V_{\mathrm{c}})= \ln\left[{\frac{V_{\mathrm{c}}^{N^{'}}}{N^{'} !}\left[\frac{\sqrt{2 m k_{B} T^{j}_{\mathrm{c}}}}{h}\right]^{3N^{'}}}\right]+\ln{\left\langle\exp \left[-\beta^{j}_{\mathrm{c}} \phi\left(\vec{q}_{1}, \cdots, \vec{q}_{N^{'}}\right)\right]\right\rangle_{\textbf{Non-Int}}}
\end{align}
and using Eq.~(\ref{eq6}), we arrive at
\begin{align}
    &\ln\mathcal{Z}(\beta^{j}_{\mathrm{c}}, N^{'}, V_{\mathrm{c}})= \ln\left[{\frac{V_{\mathrm{c}}^{N^{'}}}{N^{'} !}\left[\frac{\sqrt{2 m k_{B} T^{j}_{\mathrm{c}}}}{h}\right]^{3N^{'}}}\right]+\ln{\left\langle\exp \left[-\beta^{j}_{\mathrm{c}} \phi\left(\vec{q}_{1}, \cdots, \vec{q}_{N^{'}}\right)\right]\right\rangle_{\textbf{Non-Int}}}\nonumber\\& =\ln\left[{\frac{V_{\mathrm{c}}^{N^{'}}}{N^{'} !}\left[\frac{\sqrt{2 m k_{B} T^{j}_{{\mathrm{c}}}}}{h}\right]^{3N^{'}}}\right]+\ln{\left\langle\sum_{m=0}^{\infty} \frac{(-\beta^{j}_{\mathrm{c}})^{m}}{m !}\phi^{m}\left(\vec{q}_{1}, \cdots, \vec{q}_{N^{'}}\right)\right\rangle_{\textbf{Non-Int}}}\nonumber\\&
    =\ln\left[{\frac{V_{\mathrm{c}}^{N^{'}}}{N^{'} !}\left[\frac{\sqrt{2 m k_{B} T^{j}_{\mathrm{c}}}}{h}\right]^{3N^{'}}}\right]+\ln\left[{\sum_{m=0}^{\infty} \frac{(-\beta^{j}_{\mathrm{c}})^{m}}{m !}\left\langle\phi^{m}\left(\vec{q}_{1}, \cdots, \vec{q}_{N^{'}}\right)\right\rangle_{\textbf{Non-Int}}}\right]\nonumber\\&
    =
    \ln\left[{\frac{V_{\mathrm{c}}^{N^{'}}}{N^{'} !}\left[\frac{\sqrt{2 m k_{B} T^{j}_{\mathrm{c}}}}{h}\right]^{3N^{'}}}\right]+\ln\left[\exp\left[{\sum_{m=1}^{\infty} \frac{(-\beta^{j}_{\mathrm{c}})^{m}}{m !}\left\langle\phi^{m}\left(\vec{q}_{1}, \cdots, \vec{q}_{N^{'}}\right)\right\rangle_{\textbf{Non-Int, c}}}\right]\right]\nonumber\\&
= \ln\left[{\frac{V_{\mathrm{c}}^{N^{'}}}{N^{'} !}\left[\frac{\sqrt{2 m k_{B} T^{j}_{\mathrm{c}}}}{h}\right]^{3N^{'}}}\right]+{\sum_{m=1}^{\infty} \frac{(-\beta^{j}_{\mathrm{c}})^{m}}{m !}\left\langle\phi^{m}\left(\vec{q}_{1}, \cdots, \vec{q}_{N^{'}}\right)\right\rangle_{\textbf{Non-Int,c}}} \, .
\end{align}
Considering explicitly only the first and second term of the expansion we obtain
\begin{align}
    &\ln\mathcal{Z}(\beta^{j}_{\mathrm{c}}, N^{'}, V_{\mathrm{c}})= \ln\left[{\frac{V_{\mathrm{c}}^{N^{'}}}{N^{'} !}\left[\frac{\sqrt{2 m k_{B} T^{j}_{\mathrm{c}}}}{h}\right]^{3N^{'}}}\right]+{ \frac{(-\beta^{j}_{\mathrm{c}})}{1 !}\left\langle\phi\left(\vec{q}_{1}, \cdots, \vec{q}_{N^{'}}\right)\right\rangle_{\textbf{Non-Int,c}}}\nonumber\\& \frac{(-\beta^{j}_{\mathrm{c}})^{2}}{2 !}\left\langle\phi^{2}\left(\vec{q}_{1}, \cdots, \vec{q}_{N^{'}}\right)\right\rangle_{\textbf{Non-Int,c}}+\text{Higher order terms} \, .
\end{align}
We can substitute the internal potential, Eq.~(\ref{potential})
\begin{align}
    &\ln\mathcal{Z}(\beta^{j}_{\mathrm{c}}, N^{'}, V_{\mathrm{c}})= \ln\left[{\frac{V_{\mathrm{c}}^{N^{'}}}{N^{'} !}\left[\frac{\sqrt{2 m k_{B} T^{j}_{\mathrm{c}}}}{h}\right]^{3N^{'}}}\right]+\frac{N^{'}(N^{'}-1)}{2 V_{\mathrm{c}}} \int \mathrm{~d}^{3} \vec{r} \ \phi_{ij}({\vec{r}})\nonumber\\&+\frac{N^{'}(N^{'}-1)}{2}\left[\int \frac{\mathrm{d}^{3} \vec{r}}{V_{\mathrm{c}}} \phi_{ij}^{2}(\vec{r})-\left(\int \frac{\mathrm{d}^{3} \vec{r}}{V_{\mathrm{c}}} \phi_{ij}(\vec{r})\right)^{2}\right]+\text{Higher order terms}
\end{align}
Using mathematical induction \cite{kardar2007statistical} for higher order terms, we can write,
\begin{align}\label{eqlog}
    &\ln\mathcal{Z}(\beta^{j}_{\mathrm{c}}, N^{'}, V_{\mathrm{c}}) \nonumber\\& = \ln\left[{\frac{V_{\mathrm{c}}^{N^{'}}}{N^{'} !}\left[\frac{\sqrt{2 m k_{B} T^{j}_{\mathrm{c}}}}{h}\right]^{3N^{'}}}\right]+\sum_{m=1}^{\infty} \frac{(-\beta)^{m}}{m !} \frac{N^{'}(N^{'}-1)}{2} \int \frac{\mathrm{d}^{3} \vec{r}}{V_{\mathrm{c}}} \phi_{ij}^{{m}}(\vec{r})+\mathcal{O}\left(\frac{{N^{'}}^{3}}{V_{\mathrm{c}}^{2}}\right)\nonumber \\&= \ln\left[{\frac{V_{\mathrm{c}}^{N^{'}}}{N^{'} !}\left[\frac{\sqrt{2 m k_{B} T^{j}_{\mathrm{c}}}}{h}\right]^{3N^{'}}}\right]+\frac{N^{'}(N^{'}-1)}{2 V_{\mathrm{c}}} \int \mathrm{d}^{3} \vec{r}[\exp (-\beta \phi_{ij}(\vec{r}))-1]+\mathcal{O}\left(\frac{{N^{'}}^{3}}{V_{\mathrm{c}}^{2}}\right)
\end{align}
Terminating the expansion at order ${{N^{'}}^2}/{V_\mathrm{c}}$ and applying the exponential function on both sides of Eq.~(\ref{eqlog}), we can write 
\begin{align}\label{eqzp}
    &\mathcal{Z}(\beta^{j}_{\mathrm{c}}, N^{'}, V_{\mathrm{c}})\approx \left[{\frac{V_{\mathrm{c}}^{N^{'}}}{N^{'} !}\left[\frac{\sqrt{2 m k_{B} T^{j}_{\mathrm{c}}}}{h}\right]^{3N^{'}}}\right]\exp\left[\frac{N^{'}(N^{'}-1)}{2 V_{\mathrm{c}}} \int \mathrm{d}^{3} \vec{r}[\exp (-\beta^{j}_{\mathrm{c}} \phi_{ij}(\vec{r}))-1]\right]\, .
\end{align}
With Eq.~(\ref{eqq}), we finally obtain for the grand canonical partition function
\begin{align}
&{\mathcal{Q}(\beta^{j}_{\mathrm{c}}, \mu^{j}_{\mathrm{c}}, V_{\mathrm{c}})}=\sum_{N^{'}=0}^{\infty} \mathrm{e}^{\beta^{j}_{\mathrm{c}} \mu^{j}_{\mathrm{c}} N^{'}}\mathcal{Z}(\beta^{j}_{\mathrm{c}}, N^{'}, V_{\mathrm{c}})
\approx\sum_{N^{'}=0}^{\infty}
{\mathrm{e}^{\beta^{j}_{\mathrm{c}} \mu^{j}_{\mathrm{c}} N^{'} } \mathcal{C}_{1}(\beta^{j}_{\mathrm{c}}, N^{'}, V_{\mathrm{c}})\mathcal{C}_{2}(\beta^{j}_{\mathrm{c}}, N^{'}, V_{\mathrm{c}})}
\end{align}
in which
\begin{align}
 \mathcal{C}_{1}(\beta^{j}_{\mathrm{c}}, N^{'}, V_{\mathrm{c}})=\frac{1}{N^{'}!} \left[{{V_{\mathrm{c}}}\left[\frac{\sqrt{2 m k_{B} T^{j}_{\mathrm{c}}}}{h}\right]^{3}}\right]^{N^{'}}
\end{align}
and
\begin{align}
 \mathcal{C}_{2}(\beta^{j}_{\mathrm{c}}, N^{'}, V_{\mathrm{c}})=\exp\left[\frac{N^{'}N^{'}}{2\,V_{\mathrm{c}} }\mathcal{C}_{3}(\beta^{j}_{\mathrm{c}}, V_{\mathrm{c}})-\frac{N^{'}}{2\,V_{\mathrm{c}} }\mathcal{C}_{3}(\beta^{j}_{\mathrm{c}}, V_{\mathrm{c}})\right]\, ,
\end{align}
where
\begin{align}
\mathcal{C}_{3}(\beta^{j}_{\mathrm{c}}, V_{\mathrm{c}})= \int \mathrm{d}^{3} \vec{r}[\exp (-\beta^{j}_{\mathrm{c}} \phi_{ij}(\vec{r}))-1]\, .
\end{align}
As we will see in the following section (precisely in Eq.~(\ref{seriesexp})), we will interpret the infinite sum over $N^{'}$ as the series expansion of the exponential function. To do this, we must eliminate the dependence on $(N^{'})^2$ in $\mathcal{C}_2$. We achieve this by replacing $\mathcal{C}_{2}(\beta^{j}_{\mathrm{c}}, N^{'}, V_{\mathrm{c}})$ with $\mathcal{C}_{2}^{\left\langle N_{\mathrm{c}} \right\rangle}(\beta^{j}_{\mathrm{c}}, N^{'}, V_{\mathrm{c}})$  defined as
\begin{align}
    \mathcal{C}_{2}^{\left\langle N_\mathrm{c} \right\rangle}(\beta^{j}_{\mathrm{c}}, N^{'}, V_{\mathrm{c}}):=\exp\left[\frac{{\left\langle N_\mathrm{c} \right\rangle}N^{'}}{2\,V_{\mathrm{c}} }\mathcal{C}_{3}(\beta^{j}_{\mathrm{c}}, V_{\mathrm{c}})-\frac{{\left\langle N_\mathrm{c} \right\rangle}}{2\,V_{\mathrm{c}} }\mathcal{C}_{3}(\beta^{j}_{\mathrm{c}}, V_{\mathrm{c}})\right]
\end{align}
so that we can define ${\mathcal{Q}^{\left\langle N_\mathrm{c} \right\rangle}(\beta^{j}_{\mathrm{c}}, \mu^{j}_{\mathrm{c}}, V_{\mathrm{c}})}$ as

\begin{align}\label{}
&{\mathcal{Q}^{\left\langle N_\mathrm{c} \right\rangle}(\beta^{j}_{\mathrm{c}}, \mu^{j}_{\mathrm{c}}, V_{\mathrm{c}})}=\sum_{N^{'}=0}^{\infty} \mathrm{e}^{\beta^{j}_{\mathrm{c}} \mu^{j}_{\mathrm{c}} N^{'}}\mathcal{Z}^{\left\langle N_\mathrm{c} \right\rangle}(\beta^{j}_{\mathrm{c}}, N^{'}, V_{\mathrm{c}})\nonumber\\&
=\sum_{N^{'}=0}^{\infty}
{\mathrm{e}^{\beta^{j}_{\mathrm{c}} \mu^{j}_{\mathrm{c}} N^{'} } \mathcal{C}_{1}(\beta^{j}_{\mathrm{c}}, N^{'}, V_{\mathrm{c}})\mathcal{C}_{2}^{\left\langle N_\mathrm{c} \right\rangle}(\beta^{j}_{\mathrm{c}}, N^{'}, V_{\mathrm{c}})}\, .
\end{align}
We know that ${\mathcal{Q}(\beta^{j}_{\mathrm{c}}, \mu^{j}_{\mathrm{c}}, V_{\mathrm{c}})}$ is an infinite series of strictly non-negative terms, but it is also the grand canonical partition function that equals the phase space volume of a physical system and hence it converges to a finite number. From calculus, we know that the comparison test is a way of deducing the convergence of an infinite series. Based on the comparison test, since ${\mathcal{Q}(\beta^{j}_{\mathrm{c}}, \mu^{j}_{\mathrm{c}}, V_{\mathrm{c}})}$ converges and the following inequality holds for all sufficiently large $N^{'}$,  

\begin{align}
0<{\mathrm{e}^{\beta^{j}_{\mathrm{c}} \mu^{j}_{\mathrm{c}} N^{'} } \mathcal{C}_{1}(\beta^{j}_{\mathrm{c}}, N^{'}, V_{\mathrm{c}})\mathcal{C}_{2}^{\left\langle N_\mathrm{c} \right\rangle}(\beta^{j}_{\mathrm{c}}, N^{'}, V_{\mathrm{c}})}  <{\mathrm{e}^{\beta^{j}_{\mathrm{c}} \mu^{j}_{\mathrm{c}} N^{'} } \mathcal{C}_{1}(\beta^{j}_{\mathrm{c}}, N^{'}, V_{\mathrm{c}})\mathcal{C}_{2}(\beta^{j}_{\mathrm{c}}, N^{'}, V_{\mathrm{c}})},
\end{align}
 then the infinite series ${\mathcal{Q}^{\left\langle N_\mathrm{c} \right\rangle}(\beta^{j}_{\mathrm{c}}, \mu^{j}_{\mathrm{c}}, V_{\mathrm{c}})}$ also converges and we can write
\begin{align}
    {\mathcal{Q}(\beta^{j}_{\mathrm{c}}, \mu^{j}_{\mathrm{c}}, V_{\mathrm{c}})}=\xi\, {\mathcal{Q}^{\left\langle N_\mathrm{c} \right\rangle}(\beta^{j}_{\mathrm{c}}, \mu^{j}_{\mathrm{c}}, V_{\mathrm{c}})},\quad \xi>0
\end{align}
where $\xi$ is a scaling factor.
By doing so,
\begin{align}\label{eqqf} 
 &{\mathcal{Q}(\beta^{j}_{\mathrm{c}}, \mu^{j}_{\mathrm{c}}, V_{\mathrm{c}})}=\xi\, {\mathcal{Q}^{\left\langle N_\mathrm{c} \right\rangle}(\beta^{j}_{\mathrm{c}}, \mu^{j}_{\mathrm{c}}, V_{\mathrm{c}})}\nonumber\\&\propto \sum_{N^{'}=0}^{\infty}
{\mathrm{e}^{\beta^{j}_{\mathrm{c}} \mu^{j}_{\mathrm{c}} N^{'} } \mathcal{C}_{1}(\beta^{j}_{\mathrm{c}}, N^{'}, V_{\mathrm{c}})\exp\left[\frac{\rho N^{'}}{2 }\mathcal{C}_{3}(\beta^{j}_{\mathrm{c}}, V_{\mathrm{c}})-\frac{\rho}{2 }\mathcal{C}_{3}(\beta^{j}_{\mathrm{c}}, V_{\mathrm{c}})\right]} \, ,
\end{align}
where $\rho$ is the number density of particles inside the \textit{shell}
\begin{align}\label{dentoav}
    \rho=\frac{{\left\langle N_\mathrm{c} \right\rangle}}{V_{\mathrm{c}} } \, .
\end{align}
Obviously $\mathcal{Q}$ and $\mathcal{Q}^{{\langle N_\text{C}\rangle}}$ are related via a simple scaling factor that becomes irrelevant once the weighted distribution is normalized at fixed $\beta$.

\section{Derivation of a functional form for $p(  \mu^{j}_{\mathrm{c}},\beta^{j}_{\mathrm{c}},V_{\mathrm{c}})$}
As discussed before, to fully determine the weight function according to Eq.~(\ref{eqQ}), we need to choose $g_2\left(\mu^{j}_{\mathrm{c}},\beta^{j}_{\mathrm{c}},V_{\mathrm{c}}\right)$. By inspection of Eq.~(\ref{eqqf}), we suggest the following ansatz, $g_2\left(\mu^{j}_{\mathrm{c}},\beta^{j}_{\mathrm{c}},V_{\mathrm{c}}\right)$ can be written as
\begin{align}\label{ansatz}
g_2\left(\mu^{j}_{\mathrm{c}},\beta^{j}_{\mathrm{c}},V_{\mathrm{c}}\right)=\frac{\rho}{2 }\mathcal{C}_{3}(\beta^{j}_{\mathrm{c}}, V_{\mathrm{c}})
\end{align}
With this choice, $g_2$ can now be interpreted as a dimensionless variable with the same order of magnitude as the average number of particles in the \textit{core}. Following  Eq.~(\ref{eqQ}), the weighted distribution function can be written as
\begin{align}\label{sopho}
p\left(  \mu^{j}_{\mathrm{c}},\beta^{j}_{\mathrm{c}},V_{\mathrm{c}}\right)\nonumber&={\mathcal{Q}^{-1}\left(g_1\left(\mu^{j}_{\mathrm{c}},\beta^{j}_{\mathrm{c}},V_{\mathrm{c}}\right),g_2\left(\mu^{j}_{\mathrm{c}},\beta^{j}_{\mathrm{c}},V_{\mathrm{c}}\right)+\ln{g_2\left(\mu^{j}_{\mathrm{c}},\beta^{j}_{\mathrm{c}},V_{\mathrm{c}}\right)}\right)}\\&={\mathcal{Q}^{-1}\left(g_1\left(\mu^{j}_{\mathrm{c}},\beta^{j}_{\mathrm{c}},V_{\mathrm{c}}\right),\frac{\rho}{2 }\mathcal{C}_{3}(\beta^{j}_{\mathrm{c}}, V_{\mathrm{c}})+\ln({\frac{\rho}{2 }\mathcal{C}_{3}(\beta^{j}_{\mathrm{c}}, V_{\mathrm{c}})})\right)}\, .
\end{align}
Here, we want to remind the reader that after choosing $g_2$, $g_1$ is uniquely determined. To make our mathematical derivation easier to follow, we continue with the notation $g_2$ that is now defined, and insert Eq.~(\ref{ansatz}) in the final step. 
Ideally, we would simply strive for an analytical expression of Eq.~(\ref{sopho}). Unfortunately, this is not possible as we show in the Appendix and we need another approximation to continue. Based on the fact that $g_2$ is of the same order of magnitude as the average number of particles ${\left\langle N_\mathrm{c} \right\rangle}$ in the \textit{core} and our starting condition that the \textit{core} is small with a volume comparable to that of the particles, it is reasonable to conclude that the observed average number of particles in the \textit{core} is a small number. For $\lim{\left\langle N_\mathrm{c} \right\rangle}\rightarrow 1$, we see that $\displaystyle \lim_{{\left\langle N_\mathrm{c} \right\rangle}\rightarrow 1}{g_2\left(\mu^{j}_{\mathrm{c}},\beta^{j}_{\mathrm{c}},V_{\mathrm{c}}\right)} \approx 1$. Now, approximating $g_2\left(\mu^{j}_{\mathrm{c}},\beta^{j}_{\mathrm{c}},V_{\mathrm{c}}\right) \approx\ln\left( g_2\left(\mu^{j}_{\mathrm{c}},\beta^{j}_{\mathrm{c}},V_{\mathrm{c}}\right)\right) +1$ at the point $(g_2\left(\mu^{j}_{\mathrm{c}},\beta^{j}_{\mathrm{c}},V_{\mathrm{c}}\right))=1$, we can write 
$$g_2\left(\mu^{j}_{\mathrm{c}},\beta^{j}_{\mathrm{c}},V_{\mathrm{c}}\right)+\ln{g_2\left(\mu^{j}_{\mathrm{c}},\beta^{j}_{\mathrm{c}},V_{\mathrm{c}}\right)} \approx 2\ln{g_2\left(\mu^{j}_{\mathrm{c}},\beta^{j}_{\mathrm{c}},V_{\mathrm{c}}\right)}+1 \, .$$
To extend the range of validity for the large average number of particles in the \textit{core}, we adjust the approximation by introducing two parameters
\begin{align}\label{taylor}
g_2\left(\mu^{j}_{\mathrm{c}},\beta^{j}_{\mathrm{c}},V_{\mathrm{c}}\right)+\ln{g_2\left(\mu^{j}_{\mathrm{c}},\beta^{j}_{\mathrm{c}},V_{\mathrm{c}}\right)} \approx \lambda_1\ln{g_2\left(\mu^{j}_{\mathrm{c}},\beta^{j}_{\mathrm{c}},V_{\mathrm{c}}\right)}+\lambda_2 \, .
\end{align}
As can be seen, one would have $\lambda_1 = 2$ and $\lambda_2 = 1$ in the vicinity of the average number of particles being equal to one. We emphasize that this parameterization is neither rigorous nor unique but a reasonable choice with the correct low particle number limit that allows us to move on. With this, Eq.~(\ref{sopho}) can be reformulated to give
\begin{align}
&p\left(  \mu^{j}_{\mathrm{c}},\beta^{j}_{\mathrm{c}},V_{\mathrm{c}}\right)={\mathcal{Q}^{-1}\left(g_1\left(\mu^{j}_{\mathrm{c}},\beta^{j}_{\mathrm{c}},V_{\mathrm{c}}\right),g_2\left(\mu^{j}_{\mathrm{c}},\beta^{j}_{\mathrm{c}},V_{\mathrm{c}}\right)+\ln{g_2\left(\mu^{j}_{\mathrm{c}},\beta^{j}_{\mathrm{c}},V_{\mathrm{c}}\right)}\right)}\nonumber\\&\approx
{\mathcal{Q}^{-1}\left(g_1\left(\mu^{j}_{\mathrm{c}},\beta^{j}_{\mathrm{c}},V_{\mathrm{c}}\right),\lambda_1\ln{g_2\left(\mu^{j}_{\mathrm{c}},\beta^{j}_{\mathrm{c}},V_{\mathrm{c}}\right)}+\lambda_2\right)}\nonumber\\&\propto \left[\sum_{N^{'}=0}^{\infty}
\mathrm{e}^{\beta^{j}_{\mathrm{c}} \mu^{j}_{\mathrm{c}} N^{'} } \mathcal{C}_{1}(\beta^{j}_{\mathrm{c}}, N^{'}, V_{\mathrm{c}})
\times\exp\left[\left(\lambda_1\ln{g_2\left(\mu^{j}_{\mathrm{c}},\beta^{j}_{\mathrm{c}},V_{\mathrm{c}}\right)}+\lambda_2\right)\, N^{'}-\left(\lambda_1\ln{g_2\left(\mu^{j}_{\mathrm{c}},\beta^{j}_{\mathrm{c}},V_{\mathrm{c}}\right)}+\lambda_2\right)\right]\right]^{-1}\, .
\end{align}
The artificial chemical potential $\mu^{j}_{\mathrm{c}}$ is defined in terms of the artificial entropy $S^{j}_{\mathrm{c}}$  as
\begin{align}\label{che}
\mu^{j}_{\mathrm{c}}\equiv-T^{j}_{\mathrm{c}}\left(\frac{\partial S^{j}_{\mathrm{c}}}{\partial N^{'}}\right)_{\phi, V_{\mathrm{c}}} \, ,
\end{align}
with volume $V_{\mathrm{c}}$, internal energy $\phi$, and the number of particles $N^{'}$.
The Sackur–Tetrode equation \cite{tetrode1912chemische,sackur1911anwendung} expresses the entropy $S$ of a monoatomic ideal gas in terms of its thermodynamic states
\begin{align}\label{tero}
S^{j}_{\mathrm{c}}=N^{'} k_{B}\left[\ln \left(\frac{V_{\mathrm{c}}}{N^{'}}\left(\frac{4 \pi m \phi}{3 N^{'} h^{2}}\right)^{3 / 2}\right)+\frac{5}{2}\right]\, .
\end{align}
Inserting Eq.~(\ref{tero}) into 
Eq.~(\ref{che}) allows us to obtain the chemical potential for a monoatomic ideal gas
\begin{align}\label{dens}
\beta^{j}_{\mathrm{c}}\mu^{j}_{\mathrm{c}}=- \ln \left[\frac{V_{\mathrm{c}}}{N^{'}}\left(\frac{2 \pi m k_\text{B} T}{h^{2}}\right)^{3 / 2}\right]= \ln{\rho {{\Lambda^{j}_{\mathrm{c}}}^3}}\, ,
\end{align}
where ${{\Lambda^{j}_{\mathrm{c}}}^3}=h / \sqrt{2 \pi m k_{\mathrm{B}} T}$ is the thermal de Broglie wavelength and $h$ is Planck's constant.
For an interacting system of particles with no internal structure, i.e. no intramolecular degrees of freedom, we can write\cite{beck2006potential}
$$
\beta^{j}_{\mathrm{c}} \mu^{j}_{\mathrm{c}}=\ln {\rho \Lambda^{3}}+\beta^{j}_{\mathrm{c}} \mu^{j,{\mathrm{ex}} }_{\mathrm{c}} \, ,
$$
with the excess chemical potential $\mu^{j,{\mathrm{ex}} }_{\mathrm{c}}$, which accounts for intermolecular interactions between molecules and is given by the potential distribution theorem \cite{widom1963some,rowlinson1982molecular}. 
Equivalently, the number density can be written as 
\begin{align}
    \rho= \gamma\frac{e^{\beta^{j}_{\mathrm{c}}\mu^{j}_{\mathrm{c}}}}{{{\Lambda^{j}_{\mathrm{c}}}^3}}, \quad \gamma=\exp\,(-{\beta^{j}_{\mathrm{c}} \mu^{j,{\mathrm{ex}} }_{\mathrm{c}}})>0 \, .
\end{align}
Now we can write
\begin{align}\label{xdf}
g_2(\mu^{j}_{\mathrm{c}},\beta^{j}_{\mathrm{c}},V_{\mathrm{c}})=\gamma\,k\,{e^{\beta^{j}_{\mathrm{c}}\mu^{j}_{\mathrm{c}}}}, \quad k=\frac{ \mathcal{C}_{3}(\beta^{j}_{\mathrm{c}}, V_{\mathrm{c}})}{{2\Lambda^{j}_{\mathrm{c}}}^3}
\end{align}
and
\begin{align}\label{subln}
\lambda_1\ln{g_2(\mu^{j}_{\mathrm{c}},\beta^{j}_{\mathrm{c}},V_{\mathrm{c}})}+\lambda_2=\lambda_1\,\ln\left[{\gamma\,k\,{e^{\beta^{j}_{\mathrm{c}}\mu^{j}_{\mathrm{c}}}}}\right]+\lambda_2=\lambda_1\,{\beta^{j}_{\mathrm{c}}\mu^{j}_{\mathrm{c}}}+\lambda_1\,\ln{\gamma\,k}+\lambda_2
\end{align}
With this we can finally write
\begin{align}\label{seriesexp}
    &p\left(  \mu^{j}_{\mathrm{c}},\beta^{j}_{\mathrm{c}},V_{\mathrm{c}}\right)\nonumber\\&\propto \left[\sum_{N^{'}=0}^{\infty}
{\mathrm{e}^{\beta^{j}_{\mathrm{c}} \mu^{j}_{\mathrm{c}} N^{'} } \mathcal{C}_{1}(\beta^{j}_{\mathrm{c}}, N^{'}, V_{\mathrm{c}})\nonumber\exp\left[(\lambda_1\,{\beta^{j}_{\mathrm{c}}\mu^{j}_{\mathrm{c}}}+\lambda_1\,\ln{\gamma\,k}+\lambda_2)\, N^{'}-(\lambda_1\,{\beta^{j}_{\mathrm{c}}\mu^{j}_{\mathrm{c}}}+\lambda_1\,\ln{\gamma\,k}+\lambda_2)\right]} \right]^{-1}\nonumber\\&
\propto \left[\sum_{N^{'}=0}^{\infty}
{\mathrm{e}^{\beta^{j}_{\mathrm{c}} \mu^{j}_{\mathrm{c}} N^{'} } \mathcal{C}_{1}(\beta^{j}_{\mathrm{c}}, N^{'}, V_{\mathrm{c}})\left[\,\mathrm{e}^{\lambda_1\,\beta^{j}_{\mathrm{c}}\mu^{j}_{\mathrm{c}}}\right]^{N^{'}}\left[\mathrm{e}^{\lambda_2}\,(\gamma\,k)^{\lambda_1}\right]^{N^{'}}\left[\mathrm{e}^{\lambda_1\,\beta^{j}_{\mathrm{c}}\mu^{j}_{\mathrm{c}}}\right]^{-1}\left[\mathrm{e}^{\lambda_2}\,(\gamma\,k)^{\lambda_1}\right]^{-1}} \right]^{-1}
\nonumber\\&
\propto \left[\sum_{N^{'}=0}^{\infty}
{\mathrm{e}^{\beta^{j}_{\mathrm{c}} \mu^{j}_{\mathrm{c}} N^{'} } \frac{1}{N^{'}!} \left[{{V_{\mathrm{c}}}\left[\frac{\sqrt{2 m k_\text{B} T^{j}_{\mathrm{c}}}}{h}\right]^{3}}\right]^{N^{'}}\left[\,\mathrm{e}^{\lambda_1\,\beta^{j}_{\mathrm{c}}\mu^{j}_{\mathrm{c}}}\right]^{N^{'}}\left[\mathrm{e}^{\lambda_2}\,(\gamma\,k)^{\lambda_1}\right]^{N^{'}}\left[\mathrm{e}^{\lambda_1\,\beta^{j}_{\mathrm{c}}\mu^{j}_{\mathrm{c}}}\right]^{-1}\left[\mathrm{e}^{\lambda_2}\,(\gamma\,k)^{\lambda_1}\right]^{-1}} \right]^{-1}
\nonumber\\&
\propto \left[\left[\mathrm{e}^{\lambda_1\,\beta^{j}_{\mathrm{c}}\mu^{j}_{\mathrm{c}}}\right]^{-1}\left[\mathrm{e}^{\lambda_2}\,(\gamma\,k)^{\lambda_1}\right]^{-1}\sum_{N^{'}=0}^{\infty}\frac{1}{N^{'}!}
\left[\underbrace{{{V_{\mathrm{c}}}\left[\frac{\sqrt{2 m k_\text{B} T^{j}_{\mathrm{c}}}}{h}\right]^{3}}\,(\gamma\,k)^{\lambda_1}\mathrm{e}^{\beta^{j}_{\mathrm{c}} \mu^{j}_{\mathrm{c}}}  \mathrm{e}^{\lambda_1\,\beta^{j}_{\mathrm{c}}\mu^{j}_{\mathrm{c}}}\mathrm{e}^{\lambda_2}}_{\alpha}\right]^{N^{'}} \right]^{-1}\, .
\end{align}
Interpreting the infinite sum over $N^{'}$ as the power series of the exponential function,
\begin{align}
    \exp{\alpha}:=\sum_{N^{'}=0}^{\infty} \frac{\alpha^{N^{'}}}{N^{'} !}
\end{align}
we arrive at
\begin{align}\label{propab}
   p\left(  \mu^{j}_{\mathrm{c}},\beta^{j}_{\mathrm{c}},V_{\mathrm{c}}\right)\propto\,(\gamma\,k)^{\lambda_1}\,\mathrm{e}^{\lambda_1\,\beta^{j}_{\mathrm{c}}\mu^{j}_{\mathrm{c}}+\lambda_2}\mathrm{e}^{{-\frac{(\gamma\,k)^{\lambda_1}\,V_\mathrm{c}}{{\Lambda^{j}_{\mathrm{c}}}^{3}}}\mathrm{e}^{(\lambda_1+1)\,\beta^{j}_{\mathrm{c}} \mu^{j}_{\mathrm{c}}+\lambda_2}}\, ,
\end{align}
and with
\begin{align}
\quad\alpha_{1}=\,(\gamma\,k)^{\lambda_1},\quad
\alpha_{2}=\frac{(\gamma\,k)^{\lambda_1}\,V_\mathrm{c}}{{\Lambda^{j}_{\mathrm{c}}}^{3}},\quad \gamma>0 \,
\end{align}
a simplified notation gives
\begin{align}\label{propab2}
        p\left(  \mu^{j}_{\mathrm{c}},\beta^{j}_{\mathrm{c}},V_{\mathrm{c}}\right)\propto\alpha_{1}\,\mathrm{e}^{\lambda_1\,\beta^{j}_{\mathrm{c}}\mu^{j}_{\mathrm{c}}+\lambda_2}\mathrm{e}^{{-\alpha_{2}}\mathrm{e}^{(\lambda_1+1)\,\beta^{j}_{\mathrm{c}}\mu^{j}_{\mathrm{c}}+\lambda_2}} \, .
\end{align}
Since $\alpha_1$ and $\alpha_2$ are explicit functions of $\beta^{j}_{\mathrm{c}}$ (via $\gamma$), for the sake of simplicity, we consider all artificial baths to have a fixed temperature equal to the thermal bath surrounding the composite system, i.e. $\beta^{j}_{\mathrm{c}} = \beta$, for all $j$ and only the artificial chemical potentials are allowed to fluctuate. It is important to note that Eq.~(\ref{propab2}) is valid for any set of artificial baths with any arbitrary combination of artificial thermodynamic variables. We choose to fix the temperature for simplicity. In this case, Eq.~(\ref{eqpr}) must be adjusted to give
\begin{align}\label{} 
p\left( \{\Vec{\mathbf{q}},\Vec{\mathbf{p}}\}_{V_{\mathrm{c}}}\mid \mu,\beta,V\right) &\approx\sum^{j}p\left( \{\Vec{\mathbf{q}},\Vec{\mathbf{p}}\}_{V_{\mathrm{c}}}\mid \mu^{\prime,j}_{\mathrm{c}},V_{\mathrm{c}}\right) p\left(  \mu^{\prime,j}_{\mathrm{c}},V_{\mathrm{c}}\right) \nonumber \\&
=  \int_{\mu^\prime_{\mathrm{c}}} \text{d}\mu^\prime_{\mathrm{c}} p\left( \{\Vec{\mathbf{q}},\Vec{\mathbf{p}}\}_V\mid \mu^\prime_{\mathrm{c}},V_{\mathrm{c}}\right)p\left(  \mu^\prime_{\mathrm{c}},V_{\mathrm{c}}\right) \, .
\end{align}
In order to preserve the probability distribution as a dimensionless function, we should define  $\mu^\prime_{\mathrm{c}} :=\beta \mu_{\mathrm{c}}$. So we can equivalently write,
\begin{align}\label{} 
p\left( \{\Vec{\mathbf{q}},\Vec{\mathbf{p}}\}_{V_{\mathrm{c}}}\mid \mu,\beta,V\right) &\approx\sum^{j}p\left( \{\Vec{\mathbf{q}},\Vec{\mathbf{p}}\}_{V_{\mathrm{c}}}\mid \beta\mu^j_{\mathrm{c}},V_{\mathrm{c}}\right) p\left(  \beta\mu^j_{\mathrm{c}},V_{\mathrm{c}}\right) \nonumber \\&
= \beta  \int_{\mu_{\mathrm{c}}} \text{d}\mu_{\mathrm{c}} p\left( \{\Vec{\mathbf{q}},\Vec{\mathbf{p}}\}_V\mid \beta\mu_{\mathrm{c}},V_{\mathrm{c}}\right)p\left(  \beta\mu_{\mathrm{c}},V_{\mathrm{c}}\right)
\end{align}
in which 
\begin{align}
p\left( \{\Vec{\mathbf{q}},\Vec{\mathbf{p}}\}_{V_{\mathrm{c}}}\mid \beta\mu^j_{\mathrm{c}},V_{\mathrm{c}}\right) = p\left( \{\Vec{\mathbf{q}},\Vec{\mathbf{p}}\}_{V_{\mathrm{c}}}\mid \mu^j_{\mathrm{c}},\beta,V_{\mathrm{c}}\right) 
\end{align}
and
\begin{align}
p\left(\beta\mu^j_{\mathrm{c}},V_{\mathrm{c}}\right)=p\left(\mu^j_{\mathrm{c}},\beta,V_{\mathrm{c}}\right) \, .
\end{align}
For the remaining derivations in the fixed temperature scenario we use the following integration convention,
\begin{align}
\sum^{j}\Longrightarrow \iint_{\mu_{\mathrm{c}} \beta_{\mathrm{c}}} \text{d}\mu_{\mathrm{c}} \text{d}\beta_{\mathrm{c}}
\end{align}
which is adjusted to 
\begin{align}
\sum^{j}\Longrightarrow \beta 
 \int_{\mu_{\mathrm{c}} } \text{d}\mu_{\mathrm{c}} \, .
\end{align}
The weighted distribution function can be written as
\begin{align}
p\left(  \mu^{j}_{\mathrm{c}},\beta,V_{\mathrm{c}}\right)=\frac{\alpha_{1}\,\mathrm{e}^{\lambda_1\,\beta\mu^{j}_{\mathrm{c}}+\lambda_2}\mathrm{e}^{{-\alpha_{2}}\mathrm{e}^{(\lambda_1+1)\,\beta\mu^{j}_{\mathrm{c}}+\lambda_2}}}{\sum_{j}\alpha_{1}\,\mathrm{e}^{\lambda_1\,\beta\mu^{j}_{\mathrm{c}}+\lambda_2}\mathrm{e}^{{-\alpha_{2}}\mathrm{e}^{(\lambda_1+1)\,\beta\mu^{j}_{\mathrm{c}}+\lambda_2}}} \, ,
\end{align}
and the normalization factor of $p(  \mu^{j}_{\mathrm{c}},\beta,V_{\mathrm{c}})$ can be written as
\begin{align}
{\sum_{j}\alpha_{1}\,\mathrm{e}^{\lambda_1\,\beta\mu^{j}_{\mathrm{c}}+\lambda_2}\mathrm{e}^{{-\alpha_{2}}\mathrm{e}^{(\lambda_1+1)\,\beta\mu^{j}_{\mathrm{c}}+\lambda_2}}}   =\beta\int_{-\infty}^{\infty} \text{d}\mu_{\mathrm{c}}\, \alpha_{1}\,\mathrm{e}^{\lambda_1\,\beta\mu_{\mathrm{c}}+\lambda_2}\mathrm{e}^{{-\alpha_{2}}\mathrm{e}^{(\lambda_1+1)\,\beta\mu_{\mathrm{c}}+\lambda_2}} \, .
\end{align}
We impose a change of variables $(\lambda_1+1)\beta \mu_{\mathrm{c}}+\lambda_2=-u$, such that $\text{d}\mu_{\mathrm{c}} = -{\text{d}u}/{(\lambda_1+1)\beta}$
 \begin{align}
     &\beta\int_{-\infty}^{\infty} \text{d}\mu_{\mathrm{c}}\, \alpha_{1}\,\mathrm{e}^{\lambda_1\,\beta\mu_{\mathrm{c}}+\lambda_2}\mathrm{e}^{{-\alpha_{2}}\mathrm{e}^{(\lambda_1+1)\,\beta\mu_{\mathrm{c}}+\lambda_2}}
=\frac{\beta}{(\lambda_1+1)\,\beta}\int_{-\infty}^{\infty} \text{d}u\,\alpha_{1}\, \mathrm{e}^{-\frac{\lambda_1\,u}{\lambda_1+1}+\frac{\lambda_2}{\lambda_1+1}}\mathrm{e}^{-\alpha_{2}\mathrm{e}^{-u}}   \nonumber\\&=\frac{\beta}{(\lambda_1+1)\,\beta}\int_{-\infty}^{\infty} \text{d}u\, \alpha_{1}\,\mathrm{e}^{\frac{\lambda_2}{\lambda_1+1}} \alpha_{2}^{-\frac{\lambda_1}{\lambda_1+1}} \alpha_{2}^{\frac{\lambda_1}{\lambda_1+1}}\, \mathrm{e}^{-\frac{\lambda_1\,u}{\lambda_1+1}}\mathrm{e}^{-\alpha_{2}\mathrm{e}^{-u}}\nonumber\\&= \frac{\beta}{(\lambda_1+1)\beta}\int_{-\infty}^{\infty} \text{d}u\, \alpha_{1}\,\mathrm{e}^{\frac{\lambda_2}{\lambda_1+1}} \alpha_{2}^{-\frac{\lambda_1}{\lambda_1+1}} (\alpha_{2}\, \mathrm{e}^{-{u}})^{\frac{\lambda_1}{\lambda_1+1}}\mathrm{e}^{-\alpha_{2}\mathrm{e}^{-u}} \, .
 \end{align}
Another change of variables $\alpha_{2}\mathrm{e}^{-u}=x$, such that $\text{d}u =- {\text{d}x}/{\alpha_{2}x}$ gives
 \begin{align}
 &\frac{\beta}{(\lambda_1+1)\beta}\int_{-\infty}^{\infty} \text{d}u\, \alpha_{1}\,\mathrm{e}^{\frac{\lambda_2}{\lambda_1+1}} \alpha_{2}^{-\frac{\lambda_1}{\lambda_1+1}} (\alpha_{2}\, \mathrm{e}^{-{u}})^{\frac{\lambda_1}{\lambda_1+1}}\mathrm{e}^{-\alpha_{2}\mathrm{e}^{-u}}\nonumber\\&=\frac{\beta}{(\lambda_1+1)\beta}\alpha_{1}\,\mathrm{e}^{\frac{\lambda_2}{\lambda_1+1}}\alpha_{2}^{\frac{-2\,\lambda_1-1}{\lambda_1+1}}\int_{0}^{\infty} \text{d}x\, x^{\frac{\lambda_1}{\lambda_1+1}-1}\mathrm{e}^{-x}=\frac{\beta}{(\lambda_1+1)\beta}\alpha_{1}\,\mathrm{e}^{\frac{\lambda_2}{\lambda_1+1}}\alpha_{2}^{\frac{-2\,\lambda_1-1}{\lambda_1+1}}\Gamma(\frac{\lambda_1}{\lambda_1+1})\, ,
 \end{align}
 and we finally arrive at
 \begin{align}
        p\left(  \mu^{j}_{\mathrm{c}},\beta,V_{\mathrm{c}}\right)=\frac{\alpha_{1}\,\mathrm{e}^{\lambda_1\,\beta\mu^{j}_{\mathrm{c}}+\lambda_2}\mathrm{e}^{{-\alpha_{2}}\mathrm{e}^{(\lambda_1+1)\,\beta\mu^{j}_{\mathrm{c}}+\lambda_2}}}{\frac{\beta}{(\lambda_1+1)\beta}\alpha_{1}\,\mathrm{e}^{\frac{\lambda_2}{\lambda_1+1}}\alpha_{2}^{\frac{-2\,\lambda_1-1}{\lambda_1+1}}\Gamma(\frac{\lambda_1}{\lambda_1+1})}=\frac{{(\lambda_1+1)\beta}\,\mathrm{e}^{\lambda_1\,\beta\mu^{j}_{\mathrm{c}}+\lambda_2}\mathrm{e}^{{-\alpha_{2}}\mathrm{e}^{(\lambda_1+1)\,\beta\mu^{j}_{\mathrm{c}}+\lambda_2}}}{{\beta}\,\mathrm{e}^{\frac{\lambda_2}{\lambda_1+1}}\alpha_{2}^{\frac{-2\,\lambda_1-1}{\lambda_1+1}}\Gamma(\frac{\lambda_1}{\lambda_1+1})} \, .
\end{align}
 And we finally obtain a valid analytic expression for the weighted distribution function of the artificial thermodynamic baths with three parameters $\lambda_1, \lambda_2$ and $\alpha_2$
 \begin{align}\label{Corr}
    p\left(  \mu^{j}_{\mathrm{c}},\beta,V_{\mathrm{c}}\right)=\frac{{(\lambda_1+1)}\,\mathrm{e}^{\lambda_1\,\beta\mu^{j}_{\mathrm{c}}+\lambda_2}\mathrm{e}^{{-\alpha_{2}}\mathrm{e}^{(\lambda_1+1)\,\beta\mu^{j}_{\mathrm{c}}+\lambda_2}}}{\mathrm{e}^{\frac{\lambda_2}{\lambda_1+1}}\alpha_{2}^{\frac{-2\,\lambda_1-1}{\lambda_1+1}}\Gamma(\frac{\lambda_1}{\lambda_1+1})} \, 
\end{align}
where
\begin{align}
\alpha_{2}=\frac{(\gamma\,k)^{\lambda_1}\,V_\mathrm{c}}{{\Lambda}^{3}},\quad \gamma>0,\quad k=\frac{ \mathcal{C}_{3}(\beta, V_{\mathrm{c}})}{2{\Lambda}^{3}},\quad \lambda_1>0,\quad \Lambda=\frac{h}{\sqrt{2 \pi m k_{\mathrm{B}} T}} \nonumber
\end{align}
\begin{align}
\mathcal{C}_{3}(\beta, V_{\mathrm{c}})= \int \mathrm{d}^{3} \vec{r}[\exp (-\beta \phi_{ij}(\vec{r}))-1] \nonumber \, .
\end{align}

The purpose of this study is to capture the properties of an interacting system in an explicitly simulated \textit{core} region conditional to the thermodynamic constraints of the bath. The computational low-cost strategy was proposed in Eq.~(\ref{eqpr}). Now,
by starting from $p( N_\mathrm{c}\mid \mu,\beta,V)$ and following the same steps as in Eqns.~(\ref{eqsix}-\ref{eqpr}) the counterpart of Eq.~(\ref{eqpr}) can be written as 
\begin{align}\label{ND} 
&p\left( N_\mathrm{c}\mid \mu,\beta,V\right) \approx\sum_{j}p\left(N_\mathrm{c} \mid \mu^{j}_{\mathrm{c}},\beta^{j}_{\mathrm{c}},V_{\mathrm{c}}\right) p\left(  \mu^{j}_{\mathrm{c}},\beta^{j}_{\mathrm{c}},V_{\mathrm{c}}\right) \, .
\end{align}
In the following we will establish a workflow that allows us to accurately capture the probability distribution of the number of particles inside the \textit{core} $p( N_\mathrm{c}\mid \mu,\beta,V)$ from grand canonical simulations. For sake of simplicity, we again only allow artificial chemical potentials $\mu^{j}_{\mathrm{c}}$ to change while we keep $\beta^{j}_{\mathrm{c}}$ constant and equal to $\beta$. Eq.~(\ref{ND}) can be expanded by substituting Eq.~(\ref{Corr})
\begin{align}\label{expand}
&p\left( N_\mathrm{c}\mid \mu,\beta,V\right) \approx \beta \int_{-\infty}^{\infty}\,\text{d}\,\mu_{\mathrm{c}}\,p\left(N_\mathrm{c} \mid \mu_{\mathrm{c}},\beta,V_{\mathrm{c}}\right)\, \frac{{(\lambda_1+1)}\,\mathrm{e}^{\lambda_1\,\beta\mu_{\mathrm{c}}+\lambda_2}\mathrm{e}^{{-\alpha_{2}}\mathrm{e}^{(\lambda_1+1)\,\beta\mu_{\mathrm{c}}+\lambda_2}}}{\mathrm{e}^{\frac{\lambda_2}{\lambda_1+1}}\alpha_{2}^{\frac{-2\,\lambda_1-1}{\lambda_1+1}}\Gamma(\frac{\lambda_1}{\lambda_1+1})}  \, .
\end{align}
We use a simplified parametric notation for the right hand side of Eq.~(\ref{expand}) as follows
\begin{align}
&p\left( N_\mathrm{c}\mid \mu,\beta,V\right) \approx p\left( N_\mathrm{c}\mid \beta,V_\mathrm{c},\alpha_2,\lambda_1,\lambda_2\right) \, .
\end{align}
As we discussed in the introduction, in order to find $p( N_\mathrm{c}\mid \mu,\beta,V)$, a trivial albeit computationally expensive way is to simulate a system where a large fraction of the \textit{bath} is included in addition to the small \textit{core}. We can treat the combined system as a micro-canonical, canonical or grand canonical system under an appropriate thermodynamic constraint. Then, by using Monte Carlo simulations, we can sample the probability distribution of the number of particles in the \textit{core}. Alternatively, with our newly proposed framework that culminates in Eq.~(\ref{Corr}), we have mapped the problem --- provided the parameters $\alpha_2$, $\lambda_1$ and $\lambda_2$ have been determined appropriately --- to a computationally cheaper method. 

We will show in the following that we can determine a set of parameters that leads to accurate results for a hard-sphere model system and a Lennard--Jones fluid. To optimize these parameters, we opted to perform canonical reference calculations on a very large system. Obviously, this route only leads to a computationally cheaper method if the parameters are transferable for similar situations (e.g. \textit{core} volume, interaction strengths), such that the expensive reference simulations are rarely required. We will come back to this point in the discussion. For now, we describe how we optimized the parameters based on reference data and demonstrate that our model is indeed capable to reproduce the reference data from \textit{core}-only simulations.

To generate our reference data, we run a canonical Monte Carlo simulation for the combined system, i.e. \textit{bath+shell+core} and then sample the probability distribution of the number of particles in the \textit{core}, termed ${p_{\text{ref}}(N_\mathrm{c}\mid \mu,\beta,V)}$.
We then have to fit the parametric model $p( N_\mathrm{c}\mid \beta,V_\mathrm{c},\alpha_2,\lambda_1,\lambda_2)$ to match ${p_{\text{ref}}(N_\mathrm{c}\mid \mu,\beta,V)}$. For a given reference distribution $({p_{\text{ref}}({N^1_\mathrm{c}})}, \ldots, {p_{\text{ref}}({N^k_\mathrm{c}})})$, where $({N^1_\mathrm{c}}, \ldots, {N^k_\mathrm{c}})$ is a set of $k$ different numbers of particles inside the \textit{core}, a class of candidates for the parametric model can be defined as
\begin{align}
 &{p}({N^i_\mathrm{c}}\mid \beta,V_\mathrm{c},\theta)={p}\left( {N^i_\mathrm{c}}\mid \beta,V_\mathrm{c},\alpha_2,\lambda_1,\lambda_2\right)
  \quad \forall i=1, \ldots, k, \nonumber \\& \theta=(\alpha_2,\lambda_1,\lambda_2) \in \Theta, \quad \text{s.t.}\quad\Theta=\{(\alpha_2,\lambda_1,\lambda_2)\mid \alpha_2>0,\, \lambda_1>0 ,\,\lambda_2\in \mathbb{R}\}\, ,
\end{align}
where the structure of the function ${p}({N^i_\mathrm{c}}\mid \beta,V_\mathrm{c},\theta)$ is completely determined by Eq.~(\ref{expand}) and the $3$-dimensional parameter vector $\theta$.
A reasonable and quite intuitive way to fit the parametric model to the reference data is to use a quadratic approximation criterion. The average squared approximation error w.r.t. the observed reference
\begin{align}\label{error}
\delta(\theta):=\delta(\alpha_2,\lambda_1,\lambda_2)&=\sum_{i=1}^{k}\left[{\log p_{\text{ref}}({N^i_\mathrm{c}}\mid \mu,\beta,V)}-\log {p}({N^i_\mathrm{c}}\mid \beta,V_\mathrm{c},\theta)\right]^{2}
\end{align}
will be used to quantify the difference between the proposed model and canonical simulation results. 

Note that in the simulation section, we denote $\delta$ as the error between our proposed model, Eq.~(\ref{expand}) -- with optimized parameters $\alpha_2,\lambda_1,\lambda_2$ -- and the canonical reference calculations, whereas we show the discrepancy between the canonical simulations and the single-bath grand canonical simulations with $\Delta$ that is defined as

\begin{align}\label{errorsingle}
\Delta:=\sum_{i=1}^{k}\left[{\log p_{\text{ref}}({N^i_\mathrm{c}}\mid \mu,\beta,V)}-\log {p}_{\text{GC}}({N^i_\mathrm{c}}\mid \mu^{*}_\mathrm{c},\beta,V_\mathrm{c})\right]^{2} \, ,
\end{align}
where ${p}_{\text{GC}} ({N^i_\mathrm{c}}\mid \mu^{*}_\mathrm{c},\beta,V_\mathrm{c})$ is the single-bath
grand canonical probability distribution  with the chemical potential $\mu^{*}_\mathrm{c}$ that is defined as
\begin{equation}
\mu^{*}_\mathrm{c} := \underset{\mu^{j} \in [\mu_\text{min}, \mu_\text{max}] }{\operatorname{argmin}} \overbrace{\bigg|\left\langle {N^i_\mathrm{c}}\, p_{\text{ref}}({N^i_\mathrm{c}}\mid \mu,\beta,V)  \right\rangle- \left\langle {N^i_\mathrm{c}}\, {p}_{\text{GC}}({N^i_\mathrm{c}}\mid \mu^{j},\beta,V_\mathrm{c})\right\rangle\bigg|}^{W(\mu^{j})}\, ,
\end{equation}
where ${\operatorname{argmin}}$ stands for the value of the chemical potential in the domain of
integration $[\mu_\text{min}, \mu_\text{max}]$ that produces the minimum
value of $W(\mu^{j})$ over the the domain of
integration. Note that the bracket $\left\langle\dots\right\rangle$ indicates the average number of the particles inside the \textit{core}.

\section{Simulations}
\begin{algorithm}[h!]
\caption{Canonical Monte Carlo}\label{alg1}
\begin{algorithmic}
\State $\bullet$ Set up the particles in some initial configuration $\mathbf{r}$.
\State $\bullet$ pick a maximum displacement vector $\left(d_x, d_y, d_z\right)$.
\State $\bullet$ Initialize the Monte Carlo step counter $n_\text{MCS}=0$ and choose the maximum number of Monte Carlo steps $n_{\max }$.
\While {$n_{\mathrm{MCS}}<n_{\max }$}
    \State $\bullet$ From all $N$ particles, choose particle $i$ at random.
    \State $\bullet$ Calculate the energy of the current system $\phi_\text{ini}(\mathbf{r})$.
    \State $\bullet$ Generate three uniform random numbers $u_1, u_2, u_3$ between $-{1}$ and ${1}$.
    \State $\bullet$ Through random displacement, $\Delta \mathbf{r}=\left(u_1 d_x, u_2 d_y, u_3 d_z\right)$, \State shift atom $i$ to its new location $\mathbf{r}_i=\mathbf{r}_i+\Delta \mathbf{r}$.
    \State $\bullet$ Calculate the energy of the system after the movement $U_\text{fin}(\mathbf{r})$.
     \State $\bullet$ Find the difference in potential energy $\Delta \phi = \phi_\text{fin}-\phi_\text{ini}$,\State caused by the displacement of atom $i$,\State then the Metropolis algorithm's fundamental principle\State is that a motion is accepted or rejected based on.
    \If {$\Delta \phi \leq 0$}
        \State $\bullet$ Accept the state
    \ElsIf {$\Delta \phi>0$}
        \State $\bullet$ Generate a random number $k$ between 0 and 1.
        \If {$k<e^{-\Delta \phi /   k_\text{B} T}$}
            \State $\bullet$ Accept the state.
        \ElsIf {$k>e^{-\Delta \phi /   k_\text{B} T}$}    
        \State $\bullet$ The trial is rejected, the configuration that \State existed before the rejected move is returned: $\mathbf{r}_i=\mathbf{r}_i-\Delta \mathbf{r}$.
        \EndIf    
    \EndIf
    \State $\bullet$ Count the number of particles in the \textit{core} for positions $\mathbf{r}$, $N_\text{c}$.
    \State $\bullet$ Set
 $n_{\mathrm{MCS}}=n_{\mathrm{MCS}}+1$\
\EndWhile
\State $\bullet$
Calculate probability distribution of the number of particles in the \textit{core} based on the histogram for all $N_\text{c}$.
\end{algorithmic}
\end{algorithm}

\begin{algorithm}[h!]
\caption{Grand Canonical Monte Carlo}\label{alg2}
\begin{algorithmic}
\State $\bullet$ Initialize the Monte Carlo step counter $n_\text{MCS}=0$ and choose the maximum number of maximum Monte Carlo steps $n_{\max }$.
\While {$n_{\mathrm{MCS}}<n_{\max }$}
    \State $\bullet$ Select a uniform random number $\eta_{1}$ between 0 and 1 .
    \If{$\eta_{1}<0.5$ (particle removal attempt)}
        \State $\bullet$  Select a uniform random number $\eta_{2}$ between 0 and 1 
        \State $\bullet$ perform the Metropolis test given by
        $$
        \pi(N \rightarrow N-1)=\operatorname{Min}\left(1, \frac{\Lambda^{3} N}{V} \exp (-\beta(\mu+\phi(N-1)-\phi(N)))\right)
        $$
    \If{$\eta_{2}< \pi(N \rightarrow N-1)$ }
        \State $\bullet$ Remove the particle. Otherwise, keep the old configuration.
    \EndIf  
    \ElsIf{$\eta_{1}>0.5$ (particle insertion attempt) }
    \State $\bullet$ Select a uniform random number $\eta_{2}$ between 0 and 1 .
    \State $\bullet$ perform the Metropolis test given by
    $$
    \pi(N \rightarrow N+1)=\operatorname{Min}\left(1, \frac{V}{\Lambda^{3}(N+1)} \exp (\beta(\mu-\phi(N+1)+\phi(N)))\right)
    $$
    \If{$\eta_{2}< \pi(N \rightarrow N+1)$}
        \State $\bullet$  Add the particle. 
    \ElsIf{$\eta_{2}> \pi(N \rightarrow N+1)$}    
        \State $\bullet$ keep the old configuration.
    \EndIf
    \EndIf
    \State $\bullet$ Count the number of particles in the \textit{core}.
    \State $\bullet$ Set
 $n_{\mathrm{MCS}}=n_{\mathrm{MCS}}+1$\
\EndWhile
\State $\bullet$
Calculate probability distribution of the number of particles in the \textit{core} based on the histogram for all $N_\text{c}$.
\end{algorithmic}
\end{algorithm}

Having derived an expression for the particle number distribution in the \textit{core}, $p(N_\text{c}| \mu, \beta, V)$ (see Eq.~(\ref{expand})), we will now demonstrate that for a proper choice of parameters this expression is indeed flexible enough to accurately convert the grand canonical simulation results to those obtained from a canonical simulation with a large simulation box for an interacting system. Only if this is achieved, we can obtain accurate results for the small system of interest at the low computational cost of a grand canonical simulation of the \textit{core} system alone.

As a first check, we will test our approach for the case of a non-interacting hard-sphere model. This system has been studied before by Dixit \textit{et al.}\cite{dixit2017mini} and we compare to their results in the following. We simulate the system at the reduced density $8\rho r^{3}_\text{p}=\rho^{\prime}= 0.9$ and a fixed radius for the spherical \textit{core} or solvation shell $R = 2.2 r_\text{p}$, where $r_\text{p}$ is the radius of the hard-sphere particle. In all simulations we fix one particle in the center of the \textit{core} and count the number of additional particles in the solvation shell or \textit{core}. For the reference canonical Monte-Carlo calculations we choose a cubic box with an edge length of 
\begin{align}
    d = (8Nr^{3}_\text{p} /\rho^{\prime})^{1/3}\, ,
\end{align} 
employing periodic boundary conditions and $N=100$. 

We obtain reference canonical simulation results employing a modified version of the Metropolis algorithm~\ref{alg1}, where acceptance of a random Monte-Carlo step is only possible if there is no overlap between the particles. A similar modification is used for the grand canonical simulations following algorithm~\ref{alg2}, since no inner energy $\phi$ is associated with the configurations of the hard-sphere model.
\begin{figure}[h]
    \centering
    \includegraphics[width=0.9\textwidth]{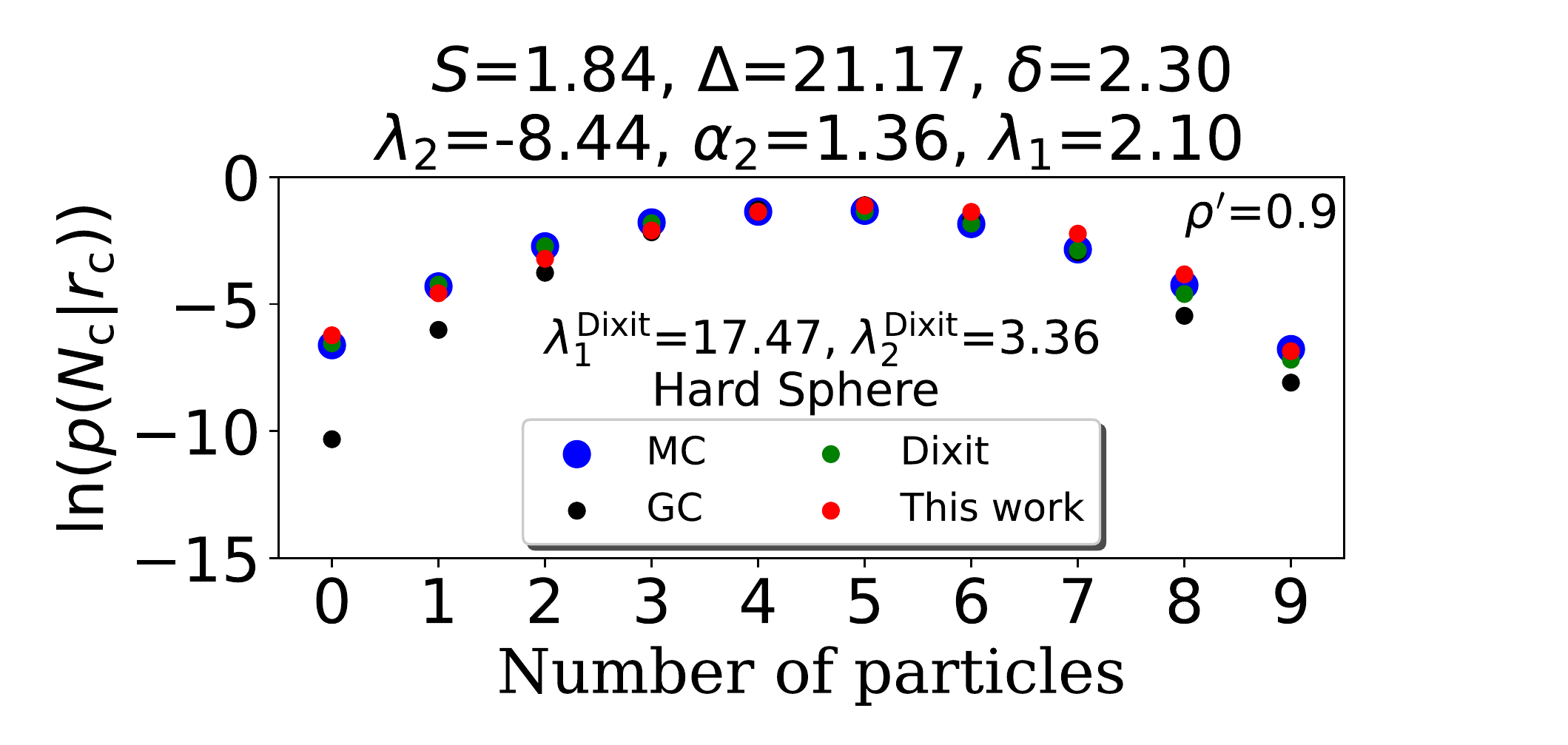}
    \caption{
    Particle number distributions for reduced density $\rho^\prime=0.9$ of a three-dimensional hard sphere  liquid. The blue dots show the results for a reference canonical Monte Carlo simulation (MC), a grand canonical Monte Carlo simulation for a single bath is shown in black (GC) and results for the improved grand canonical description with coupling to multiple baths is shown in green (Dixit, Ref.~\citenum{dixit2017mini}) and red (this work). The parameters above the graph show the optimized parameters for the model suggested in this work, the corresponding entropy is denoted by $S$ and $\Delta$ denotes the discrepancy between the reference calculation and the single-bath grand canonical simulation according to Eq.~(\ref{errorsingle}). The error between our proposed model, Eq.~(\ref{expand}) -- with optimized parameters $\alpha_2,\lambda_1,\lambda_2$ -- and the  reference calculations has been denoted by $\delta$ according to Eq.~(\ref{error}) .}
    \label{hardsphere}
\end{figure}

In Figure~\ref{hardsphere} we show the particle number distribution obtained from the canonical reference calculations (blue dots), the standard grand canonical simulation assuming a single bath (black dots), the corrected prediction according to Dixit \textit{et al.} (green dots) and the corrected prediction according to Eq.~(\ref{expand}) (red dots). 
We implemented a numerical integration scheme to approximate the integral in Eq.~(\ref{expand}). To do this, we had to transform the domain of integration in $\mu_\text{c}$ from an indefinite to a definite one, $\mu^{j} \in [\mu_\text{min}, \mu_\text{max}]$. The integrand has to be evaluated at a finite set of integration points i.e. chemical potentials and a weighted sum of these values is employed to approximate the integral via the trapezoidal rule. We carefully checked convergence with respect to the bounds of the integral and the intervals applied for the numerical integration for each individual calculation. Note that the integration points and weights rely on the specific scheme employed and the accuracy demanded from the calculation.
As can be seen in Eq.~(\ref{expand}), the integrand comprises two parts, the first part is $p\left(N_\text{c}\mid \mu^{j}_\text{c},\beta,V_\text{c}\right)$, which is evaluated with Metropolis grand canonical Monte Carlo sampling at each integration point, Algorithm \ref{alg2}, and the second part is the weight function that can be analytically evaluated as a function of the parameters $\theta = (\alpha_2, \lambda_1,\lambda_2)$.
Now, $\alpha_2$, $\lambda_1$ and $\lambda_2$ can be determined such that they minimize the error between $\log {p_{\text{ref}}({N^i_\mathrm{c}}\mid \mu,\beta,V)}$ and $\log {p}({N^i_\mathrm{c}}\mid \beta,V_\mathrm{c},\alpha_2,\lambda_1,\lambda_2)$, Eq.~(\ref{error}).
We employed a similar strategy to obtain optimal parameters for the model of Dixit \textit{et al.}\cite{dixit2017mini} (green dots).

While there is a discrepancy of $\Delta=21.17$, Eq.~(\ref{errorsingle}), between the single bath grand canonical result and the reference simulations, both optimized multi-bath descriptions (green and red dots) are perfectly capable to reproduce the reference results, reducing the squared error to only $\delta=2.30$. That our approach produces roughly the same result as the one of Dixit \textit{et al.} is to be expected since for the choice $\alpha_2=1.0$ the functional form of our model equals that of Dixit \textit{et al.}\cite{dixit2017mini} as we show in the following.\\
Their weighted distribution function reads
\begin{align}\label{dixit}
P(\mu)=\frac{\mathrm{e}^{\left(-\lambda_1^{\prime}\,\mathrm{e}^{-\mu^\prime}-\lambda_2^{\prime} \mu^\prime\right)}} \Gamma\left(\lambda_2^{\prime}\right) \lambda_1^{{\prime}^{-\lambda_2^{\prime}}}=\frac{\mathrm{e}^{-\lambda_1^{\prime} \mathrm{e}^{-\mu^\prime}}\mathrm{e}^{-\lambda_2^{\prime} \mu^\prime}}{\Gamma\left(\lambda_2^{\prime}\right) \lambda_1^{{\prime}^{-\lambda_2^{\prime}}}}
\end{align}
in which $\mu^\prime$ is the chemical potential in units of $\beta$. Writing our proposed model, Eq. (\ref{Corr}), in the units of  Dixit \textit{et al.} ($\mu^\prime=\beta \mu_c^j$) we obtain

\begin{align}
p\left(  \mu^\prime,\beta,V_{\mathrm{c}}\right)
=\frac{{(\lambda_1+1)}\,\mathrm{e}^{\lambda_1\,\mu^\prime}\mathrm{e}^{\lambda_2}\mathrm{e}^{{-\alpha_{2}}\mathrm{e}^{(\lambda_1+1)\,\mu^\prime}\mathrm{e}^{\lambda_2}}}{\mathrm{e}^{\frac{\lambda_2}{\lambda_1+1}}\alpha_{2}^{\frac{-2\,\lambda_1-1}{\lambda_1+1}}\Gamma(\frac{\lambda_1}{\lambda_1+1})} \, .
\end{align}
Setting $\alpha_2=1.0$ considerably simplifies  the expression to
\begin{align}
&p\left(  \mu^\prime,\beta,V_{\mathrm{c}}\right)
=\frac{{(\lambda_1+1)}\mathrm{e}^{\lambda_2}\,\mathrm{e}^{-\mathrm{e}^{\lambda_2}\mathrm{e}^{(\lambda_1+1)\,\mu^\prime}}\,\mathrm{e}^{\lambda_1\,\mu^\prime}}{\Gamma(\frac{\lambda_1}{\lambda_1+1})\,\mathrm{e}^{\frac{\lambda_2}{\lambda_1+1}}}\, .
\end{align}
With the following change of variables
$$
\lambda_1+1=A \quad\text{and} \quad \mathrm{e}^{\lambda_2} = B \, ,
$$
we can write

\begin{align}
&p\left(  \mu^{j}_{\mathrm{c}},\beta,V_{\mathrm{c}}\right)=\frac{A\,B\,\mathrm{e}^{-B\mathrm{e}^{A\,\mu^\prime}}\,\mathrm{e}^{(A-1)\,\mu^\prime}}{\Gamma(\frac{A-1}{A})\,B^\frac{1}{A}} \, 
\end{align}
and arrive at the same functional form as in Eq. (\ref{dixit}). However, it is not possible to map our set of parameters to those of Dixit \textit{et al.}
In the following we will only present results obtained with our model and note here that the model in Ref. \citenum{dixit2017mini} is capable to provide results of similar quality also for an interacting system. In practice, however, we found it slightly easier to optimize the parameters of our model due to well-defined boundaries and a smaller range of values.

We now challenge our approach with an interacting Lennard--Jones fluid. For such a system, we can modulate the reduced interaction strength $\epsilon^{\prime}$ (and the distance at the potential minimum $\sigma$) in addition to the reduced density $\rho^{\prime}$. We again obtain the optimal parameters for our model by minimizing the error function $\delta$ defined in Eq.~(\ref{error}).
To understand how different choices of the parameters affect the model performance and in order to provide a reliable initial guess for subsequently applied standard numerical optimization algorithms, we prepared contour plots of the sum-of-squares error function $\delta$ for a 2D-scan of the $\lambda_1$ and $\alpha_2$ parameters, keeping $\lambda_2$ fixed. We varied $\lambda_1$ from 0 to 6 in intervals of 0.1, and $\alpha_2$ from 0 to 2 in intervals of 0.1. Contour plots where then created for values of $-100 \leq \lambda_2 \leq 10$ in steps of 10.

\begin{figure}[h]
    \centering
\includegraphics[width=1.0\textwidth]{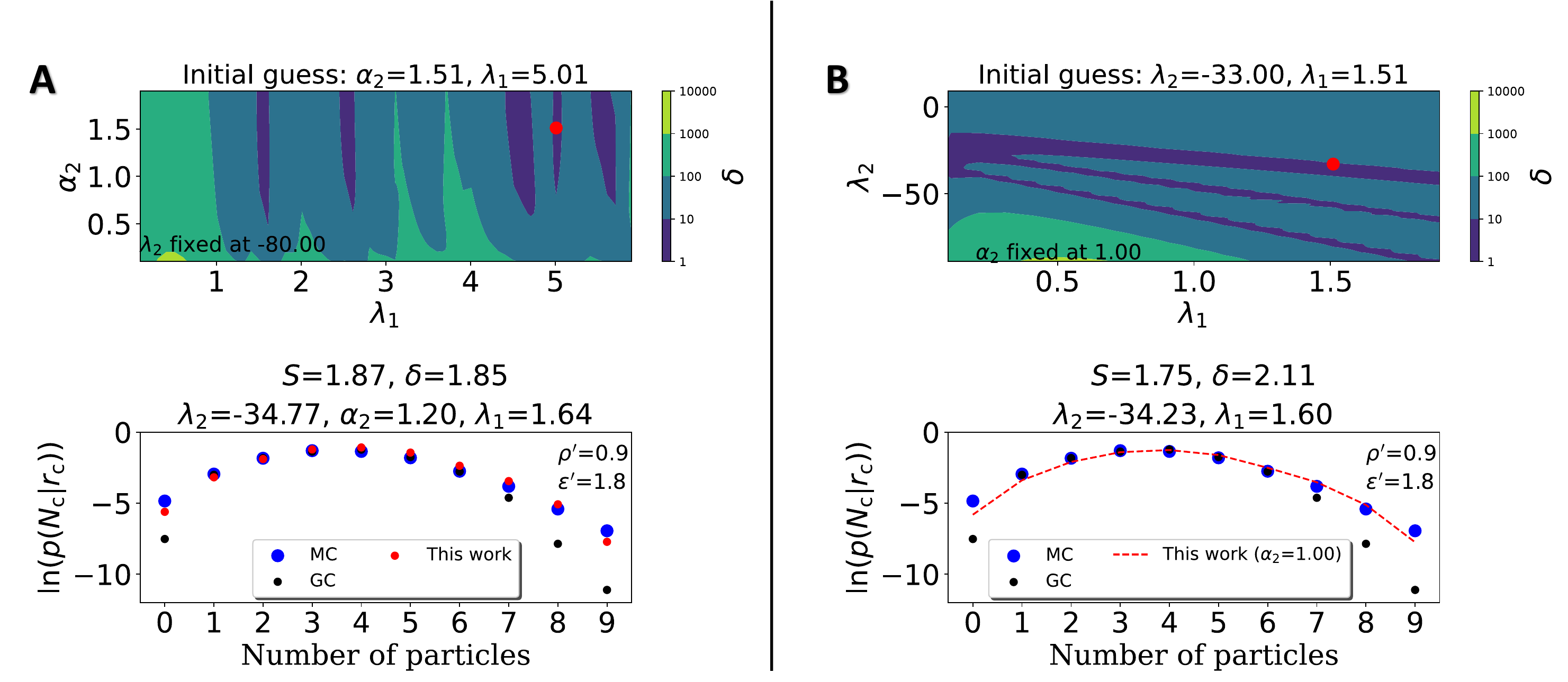}
    \caption{The upper part of panel \textbf{A} shows the contour plot of the sum-of-squares error function $\delta$ according to Eq.~(\ref{error}), for a 2D-scan of $\lambda_1$ varied between 0 and 6 in intervals of 0.1 and $\alpha_2$ varied between 0 and 2 in intervals of 0.1, keeping $\lambda_2=-80$ fixed for a constant well depth $\epsilon^{\prime} = 1.8$ of a Lennard--Jones liquid with a reduced density of $\rho^\prime=0.9$. 
    In the upper part of panel \textbf{B} we show the contour plot of the sum-of-squares error function $\delta$, for a 2D-scan of $\lambda_1$ varied between 0 and 2 in intervals of 0.1, and $\lambda_2$ varied between -100 and 10 in intervals of 10, keeping $\alpha_2=1$  fixed for the same Lennard--Jones liquid.
    The lower panels are the corresponding particle number distributions. The blue dots show the results for a reference canonical Monte Carlo simulation (MC), a grand canonical Monte Carlo simulation for a single bath is shown in black (GC) and results for the improved grand canonical description with coupling to multiple baths are shown in red dots (panel \textbf{A}, three-parameter optimization) and the red dashed line  (panel \textbf{B}, two-parameter optimization, $\alpha_2=1.0$). The parameters above each graph show the optimized parameters for the model suggested in this work for the initial guesses $(\alpha_2=1.51, \lambda_1=5.01,\lambda_2=-80)$ and $(\alpha_2=1, \lambda_1=1.51,\lambda_2=-33)$.  The corresponding entropy is denoted by $S$ and $\delta$ denotes the error between our proposed model, Eq.~(\ref{expand}) -- with optimized parameters $\alpha_2,\lambda_1,\lambda_2$ -- and the  reference calculations according to Eq.~(\ref{error}).}
    \label{initial_guess}
\end{figure}

The upper part of panel~\textbf{A}  in Figure~\ref{initial_guess} shows an exemplary result of such calculations for  $\lambda_2=-80.0$ for a reduced density of $\rho^{\prime}=0.9$ and $\epsilon^{\prime}=1.8$. The parameter $\sigma$ was fixed to a value of one for all simulations in the current work.
Similarly, the upper part of panel~\textbf{B} in  Figure~\ref{initial_guess} shows a contour plot  of the sum-of-squares error function $\delta$ for a 2D-scan of the $\lambda_1$ and $\lambda_2$ parameters, keeping $\alpha_2= 1$ fixed for the same set of $\rho^{\prime}$ and $\epsilon^{\prime}$. Keeping $\alpha_2$ fixed at 1.0 is motivated by the analogy to the approach of Dixit \textit{et al.}\cite{dixit2017mini} and by the fact that the optimized value for $\alpha_2$ is close to 1.0 for all systems studied in this work (\textit{vide infra}). We varied $\lambda_1$ from 0 to 2 in intervals of 0.1, and $\lambda_2$ from -100 to 10 in intervals of 10. We observe structured contours and substantially different local minima for both scans (red dots).  
These local minima then served as initial guesses for the numerical parameter optimization for which we chose a conjugate gradient algorithm with numerical gradients. It is worth mentioning that for panels \textbf{A} and \textbf{B} in Figure~\ref{initial_guess}, we implemented the minimization algorithm for a three-parameter optimization problem and a two-parameter (keeping $\alpha_2= 1$, fixed) optimization problem, respectively. The lower panels of Figure~\ref{initial_guess} show the particle number distribution function obtained for the optimized set of parameters. We see that we arrive at the approximately same set of corresponding optimized parameters within numerical precision of the optimization algorithm and hence quasi-identical particle number distribution functions and residual errors $\delta$ (lower graphs of both panels in Figure~\ref{initial_guess}) no matter if we keep $\alpha_2$ fixed at 1.0 or optimize it. We can further characterize the particle number distributions with an entropy
\begin{align}
    S = - \sum_{i=1}^k p \ln p 
\end{align}
and find approximately identical entropies for both simulations. If we consider the reference distribution to be the accurate equilibrium distribution, then we can interpret $\delta$ as the error between the distribution of our proposed model and the equilibrium distribution. A smaller $\delta$ can be interpreted as how much our model distribution converges toward the equilibrium distribution. However, from a statistical mechanics perspective, the entropy attributed to the equilibrium distribution should be at its maximum and any slight difference from the equilibrium distribution should lead to a decrease in entropy. 
One interesting observation is that at panel \textbf{A} we have higher entropy compared to panel \textbf{B}, which is in good agreement with observing a lower $\delta$ for panel \textbf{A} compared to panel \textbf{B}. Despite these clearly rugged hypersurfaces it is obviously possible to optimize the three parameters or two parameters (keeping $\alpha_2= 1$) of our model with standard optimization schemes and arrive at a global minimum if a suitable initial guess is provided.
As expected, the excellent agreement between the reference simulations and our model can not be obtained if only a single bath is assumed for the grand canonical.

We investigate the performance of these different models further in the following for different reduced densities $\rho^{'}$ and interaction strengths $\epsilon^{\prime}$.

\begin{figure}[h!]
    \centering
\includegraphics[width=0.48\textwidth]{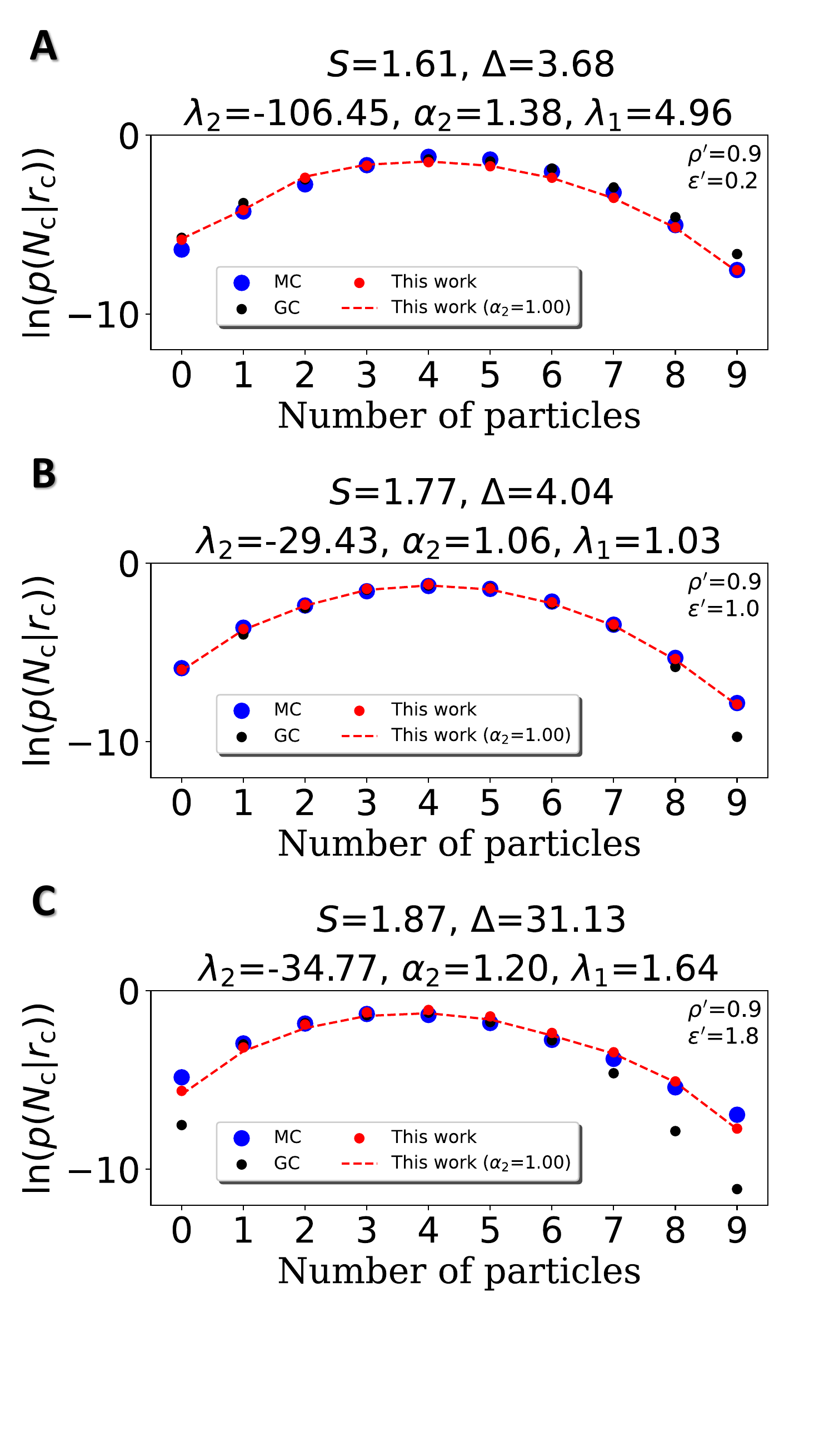}
    \caption{Particle number distributions for different reduced potential well depths $\epsilon^{\prime} = 0.2, 1.0$ and $1.8$ (panels \textbf{A}, \textbf{B} and \textbf{C}, respectively) and constant reduced density $\rho^\prime=0.9$ of a Lennard--Jones liquid. The blue dots show the results of a reference canonical Monte Carlo simulation (MC), a grand canonical Monte Carlo simulation for a single bath is shown in black (GC) and results for the improved grand canonical description with coupling to multiple baths are shown in red dots and red dashed line (this work with three parameters optimization and this work with two parameters optimization keeping $\alpha_2=1$ fixed, respectively). The optimized parameters for the model suggested in this work (with three parameters optimization) are shown above each panel along with the corresponding entropy $S$ and $\Delta$ which denotes the discrepancy between the reference calculation and the single-bath grand canonical simulation according to Eq.~(\ref{errorsingle}).}
    \label{fixdens}
\end{figure}

Figure~\ref{fixdens} shows the probability distribution of the number of particles in the solvation shell for different reduced depths of the Lennard--Jones potential well $\epsilon^{\prime}=(0.2, 1.0, 1.8)$, and a fixed reduced density $\rho^\prime=0.9$.  We observe that with increasing $\epsilon^{\prime}$, the discrepancy $\Delta$ between the single-bath grand canonical and the canonical Monte Carlo simulations becomes more pronounced. In all cases, however, we are able to identify a set of parameters for which the multi-bath model we propose here results in excellent agreement with the reference simulation.

\begin{figure}[h!]
    \centering
\includegraphics[width=0.96\textwidth]{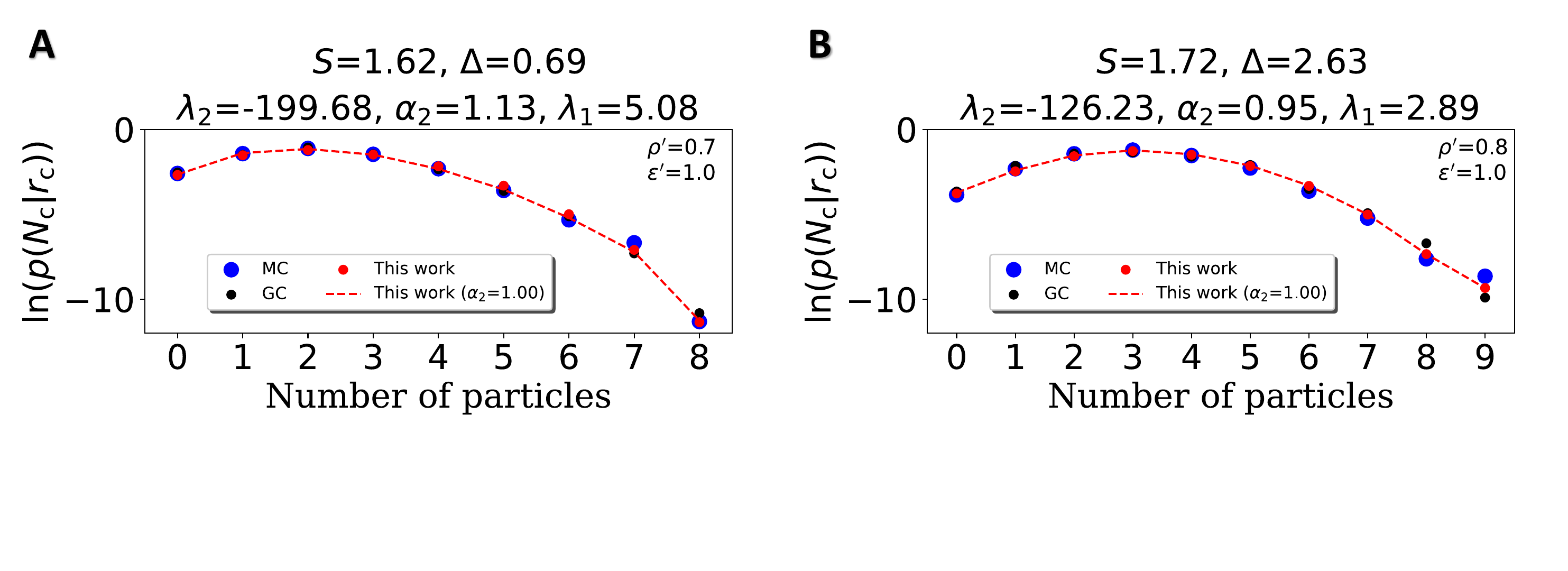}
\caption{Particle number distributions for constant reduced well depth $\epsilon^{\prime} = 1.0$ of a Lennard--Jones liquid for a reduced density of $\rho^\prime=0.7$ (panel \textbf{A}) and $\rho^\prime=0.8$ (panel \textbf{B}). The blue dots show the results for a reference canonical Monte Carlo simulation (MC), a grand canonical Monte Carlo simulation for a single bath is shown in black (GC) and results for the improved grand canonical description with coupling to multiple baths are shown in red dots and red dashed line (this work with three parameters optimization and this work with two parameters optimization keeping $\alpha_2=1$ fixed, respectively). The optimized parameters for the model suggested in this work are shown above the graphs, along with the corresponding entropy $S$ and $\Delta$, which denotes the discrepancy between the reference calculation and the single-bath grand canonical simulation according to Eq.~(\ref{errorsingle}).}
    \label{fixeps}
\end{figure}

In Figure~\ref{fixeps} we show the probability distribution of the number of particles in the solvation shell, for different reduced densities $\rho^\prime = (0.7, 0.8)$, and a fixed reduced depth of the Lennard--Jones potential well $\epsilon^{\prime}=1.0$. We observe that the smaller $\rho^\prime$, the smaller the discrepancies between the single-bath grand canonical and the canonical Monte Carlo reference simulation.

 Figures~\ref{fixdens} and \ref{fixeps} illustrate how the density and the potential well depth affect the difference between the solvent-shell particle number distribution of the single-bath grand canonical simulation and the canonical reference simulations.
 The single-bath grand canonical description for a small solvation shell is increasingly inaccurate for higher densities and more attractive potentials suggesting the average pairwise internal interaction between two particles to be the relevant property.
For a system in thermal equilibrium with a heat bath at a fixed temperature, we can calculate the classical average inter-particle distance $l$ as
\begin{align}
l \propto  \rho^{-\frac{1}{3}} \, .
 \end{align}
Since the magnitude of a pairwise force $F_r$ experienced by a particle can be defined as the negative of the derivative of the internal potential energy function with respect to the inter-particle distance $r$, the average pairwise interaction $\left\langle F_{r} \right\rangle$ for a system at thermal equilibrium can be calculated as
\begin{align*}
\left\langle F_{r} \right\rangle=
F_{r}(l)=- \frac{\partial \phi_{\text{LJ}}}{\partial{r}}\bigg|_{r=l}=-\zeta\bigg|_{r=l}\, ,
\end{align*}
where $\zeta$ is the slope of the internal potential energy curve. 
\begin{figure}[h!]
    \centering
\includegraphics[width=0.6\textwidth]{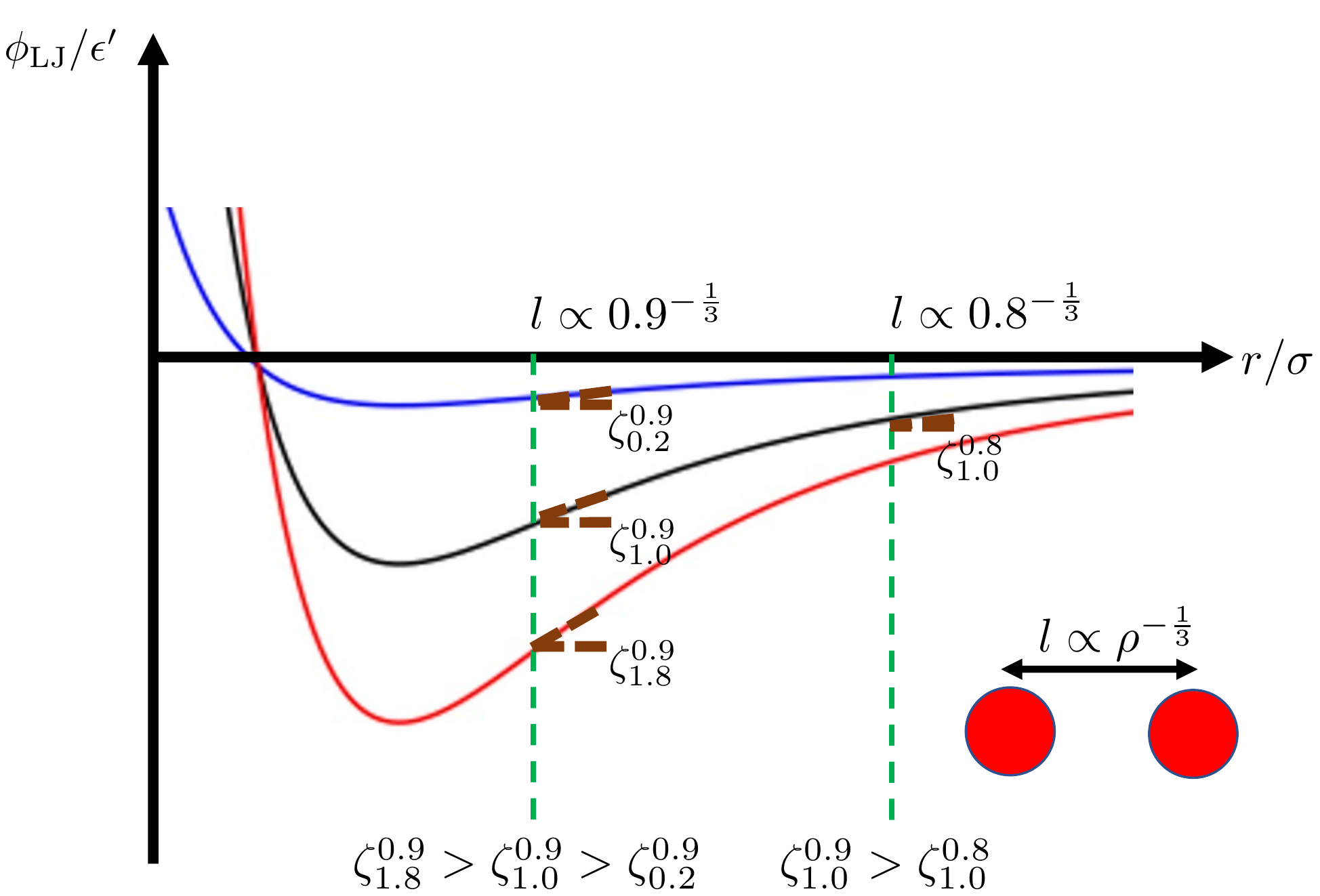}
    \caption{The pairwise Lennard--Jones potential energy $\phi_{\text{LJ}}$ as a function of the distance $r$ between a pair of particles for different reduced potential well depth $\epsilon^{\prime} = 0.2, 1.0$ and $1.8$ (curves blue, black and red, respectively) and constant $\sigma=1$ that is the distance at which the particle-particle potential energy is zero. $l$ is the classical inter-particle spacing that is controlled by the density. $\zeta_i^j$ is the slope of the internal potential energy curve at a specific density $j$ and a specific depth of potential well $i$, such that its negative value equals the force on the particle. }
    \label{free path}
\end{figure}
Figure~\ref{free path} shows the dependence of the average pairwise interaction in a Lennard--Jones system with density and depth of potential well. For a fixed density, increasing the depth of the potential well increases the magnitude of the average pairwise interaction, and for a fixed depth of the potential well, increasing the density increases the average pairwise interaction. In light of the previous observations we find that weaker pairwise interactions lead to a smaller discrepancy between the single-bath grand canonical simulation and canonical
reference calculation, making the former a more suitable approximation in these cases.

\newpage

\section{Discussion and outlook}

We present a model for the simulation of the thermodynamic properties of a small explicit \textit{core} region --- or solvation shell --- from grand canonical Monte Carlo simulations by coupling to an ensemble of thermodynamic baths rather than a single bath.
To achieve this, we introduced a weight distribution function embedded in the definition of the marginal probability distribution. We then considered a pairwise interaction between the particles using Lennard--Jones' internal potential function and derived an analytical functional form for the weight distribution. In an initial proof-of-principle numerical experiment, we showed that our model is capable to  correctly predict the particle number distribution of a non-interacting system of hard-sphere particles. Then, we challenged our model with an interacting Lennard--Jones fluid for different interaction strengths $\epsilon^{\prime}$ and reduced densities $\rho^\prime$. The simulation results demonstrate that our proposed model allows us to simulate the particle number distribution of an interacting system in the framework of computationally economic grand canonical simulations.

We note that we discarded higher-order terms in Eq.~(\ref{eqlog}) in order to arrive at a suitable analytic form for the canonical partition function. Potentially, including more terms in Eq.~(\ref{eqlog}) will result in a mathematically well-behaved model and make the optimization process faster. Such an analysis is left for future work if the current model proves to be insufficient or not flexible enough for more complicated interactions.

For the numerical integration of the newly derived expression for the particle number distribution in the solvation shell, Eq.~(\ref{expand}), we had to transform the indefinite integral over the chemical potential into a definite one. Currently, we determine both the limits of the integral and the number of integration points based on a manual convergence check. In an ongoing project, we attempt to automate this process and reduce the computational effort of this step.

An obvious drawback of our current study is certainly the need to optimize the three (or two) parameters of our model, which requires reference data that we generate from a canonical Monte Carlo simulation for the entire system which is usually rather computationally expensive. 
In upcoming research, we will investigate the transferability of these optimized parameters, i.e. if a parameter set optimized for a given reference system can be used for a slightly modified system (e.g. conformational changes) if critical properties like the size of the solvation shell remain constant. If so, a single reference calculation would allow us to perform the calculation of thermodynamical properties or even reactivity studies in this new grand canonical framework.
In addition, we will explore if a parameter optimization based on the principle of maximum entropy is a valid alternative that eradicates the need for reference data.

Last but not least, it is remarkable that Eq.~(\ref{expand})  provides a discrete-state partition function to represent the distribution of the number of particles inside the \textit{core}. Quasi-chemical theories are based on this feature \cite{beck2006potential}, although a discrete-state partition function is the primary obstacle in the construction of a quasi-chemical model. This conceptual resemblance inspired us to think about how we can link our proposed model with quasi-chemical theories. Since we incorporate the excess chemical potential inside the definition of $\gamma$ whereas in quasi-chemical theories, the chemical potential of the solvated species is usually expressed in terms of ideal and excess contributions \cite{pratt1999quasi}, we consider this to be a good starting point for linking the two models.

Most importantly, in future work we will explore how appropriate our grand canonical model describes more complicated interactions provided by advanced force fields or the full quantum-chemical Hamiltonian. In an ongoing project, we therefore calculate thermodynamic properties of solvated systems and analyze the computational advantage of our multi-bath grand canonical approach compared to a standard canonical simulations or other alternatives. In addition to solvation studies and the development of novel solvation models, we assume that our model will prove useful for the simulation of properties in metal-organic frameworks and other nano-confined structures, where the rather small system of interest (e.g. an active site) is in indirect contact with its environment.

\section{Appendix}\label{append}
\subsection{Chain rule for random variables}\label{apb}
For a set of random variables with indices $X_{1}, \ldots, X_{n}$, we can use the conditional probability definition to determine the value of each component of the joint distribution
\begin{align}
  \mathrm{p}\left(X_{n}, \ldots, X_{1}\right)=\mathrm{p}\left(X_{n} \mid X_{n-1}, \ldots, X_{1}\right) \cdot \mathrm{p}\left(X_{n-1}, \ldots, X_{1}\right)  \, .
\end{align}
The chain rule for random variables is obtained by applying the aforementioned rule for each last term
\begin{align}
   \mathrm{p}\left(\bigcap_{k=1}^{n} X_{k}\right)=\prod_{k=1}^{n} \mathrm{p}\left(X_{k} \mid \bigcap_{j=1}^{k-1} X_{j}\right) \, .
\end{align}
With three variables the chain rule produces the following product of conditional probabilities
\begin{align}
\mathrm{p}\left( X_{3}, X_{2}, X_{1}\right) =\mathrm{p}\left(X_{3} \mid X_{2}, X_{1}\right) \cdot \mathrm{p}\left(X_{2} \mid  X_{1}\right) \cdot \mathrm{p}\left( X_{1}\right)\, .
\end{align}

\subsection{Canonical partition function of the system of interacting particles}\label{canon}
The canonical partition function for a system of interacting particles with a Lennard--Jones internal interaction potential can be written as
\begin{align}\label{equn}
   &\mathcal{Z}(\beta^{j}_{\mathrm{c}}, N^{'}, V_{\mathrm{c}})=\frac{1}{N^{'} !} \int \prod_{i=1}^{N^{'}}\left(\frac{\mathrm{d}^{3} \vec{p}_{i} \mathrm{~d}^{3} \vec{q}_{i}}{h^{3}}\right) \exp \left[-\beta^{j}_{\mathrm{c}} \sum_{i} \frac{\vec{{p}_{i}}^{2}}{2 m}\right] \exp \left[-\beta^{j}_{\mathrm{c}} \phi\left(\vec{q}_{1}, \cdots, \vec{q}_{N^{'}}\right)\right]\nonumber\\&
=\frac{1}{N^{'} !} \frac{1}{h^{3N^{'}}} \int \prod_{i=1}^{N^{'}}{\mathrm{d}^{3} \vec{p}_{i} }\exp \left[-\beta^{j}_{\mathrm{c}} \sum_{i} \frac{\vec{{p}_{i}}^{2}}{2 m}\right] \int \prod_{i=1}^{N^{'}}{ \mathrm{~d}^{3} \vec{q}_{i}}\exp \left[-\beta^{j}_{\mathrm{c}} \phi\left(\vec{q}_{1}, \cdots, \vec{q}_{N^{'}}\right)\right]\nonumber\\&
=\underbrace{\frac{1}{N^{'} !} \frac{1}{h^{3N^{'}}} \left[\int {\mathrm{d} {p}_{i} }\exp \frac{{{p}_{i}}^{2}}{2 m k_{B} T^{j}_{\mathrm{c}}}\right]^{3N^{'}} }_{\frac{1}{N !}\left[\frac{\sqrt{2 m k_{B} T^{j}_{\mathrm{c}}}}{h}\right]^{3N^{'}}}\int \prod_{i=1}^{N^{'}}{ \mathrm{~d}^{3} \vec{q}_{i}}\exp \left[-\beta^{j}_{\mathrm{c}} \phi\left(\vec{q}_{1}, \cdots, \vec{q}_{N^{'}}\right)\right]\nonumber\\&
={\frac{1}{N^{'} !}\left[\frac{\sqrt{2 m k_{B} T^{j}_{\mathrm{c}}}}{h}\right]^{3N^{'}}}\int \prod_{i=1}^{N^{'}}{ \mathrm{~d}^{3} \vec{q}_{i}}\exp \left[-\beta^{j}_{\mathrm{c}} \phi\left(\vec{q}_{1}, \cdots, \vec{q}_{N^{'}}\right)\right] \nonumber\\&
\end{align}
We can now multiply Eq.~(\ref{equn}) with the sum of the probability density function evaluated over the domain of the entire phase space of a non-interacting system of particles, which yields 1:

\begin{align}\label{eqwith} 
&\mathcal{Z}(\beta^{j}_{\mathrm{c}}, N^{'}, V_{\mathrm{c}})={\frac{1}{N^{'} !}\left[\frac{\sqrt{2 m k_{B} T^{j}_{\mathrm{c}}}}{h}\right]^{3N^{'}}}\int \prod_{i=1}^{N^{'}}{ \mathrm{~d}^{3} \vec{q}_{i}}\exp \left[-\beta^{j}_{\mathrm{c}} \phi\left(\vec{q}_{1}, \cdots, \vec{q}_{N^{'}}\right)\right]\times 1 \nonumber\\& 
={\frac{1}{N^{'} !}\left[\frac{\sqrt{2 m k_{B} T^{j}_{\mathrm{c}}}}{h}\right]^{3N^{'}}}\int \prod_{i=1}^{N^{'}}{ \mathrm{~d}^{3} \vec{q}_{i}}\exp \left[-\beta^{j}_{\mathrm{c}} \phi\left(\vec{q}_{1}, \cdots, \vec{q}_{N^{'}}\right)\right]\nonumber\\& \times \int \prod_{i=1}^{N^{'}}\left(\frac{\mathrm{d}^{3} \vec{p}_{i} \mathrm{~d}^{3} \vec{q}_{i}}{h^{3}}\right)  \frac{1}{\mathcal{Z_{\textbf{Non-Int}}}(\beta^{j}_{\mathrm{c}}, N^{'}, V_{\mathrm{c}})} \exp \left[-\beta^{j}_{\mathrm{c}} \sum_{i=1}^{N^{'}} \frac{p_{i}^{2}}{2 m}\right]
\end{align}
In Eq.~(\ref{eqwith}), the integrand can be separated into two distinct parts, one over the momentum and one over the position. The position integrand, which corrsponds to a non-interacting system, has an exponential term that expresses the total kinetic energy and can be merged with the integrand of the total internal potential function of the interacting system. The result is a double-integral that can be considered as the expectation value of the internal potential function multiplied by $\beta$. It is also worth mentioning that this expectation value has been calculated based on the probability density function of a non-interacting system of particles.

\begin{align}
&\mathcal{Z}(\beta^{j}_{\mathrm{c}}, N^{'}, V_{\mathrm{c}})={\frac{1}{N^{'} !}\left[\frac{\sqrt{2 m k_{B} T^{j}_{\mathrm{c}}}}{h}\right]^{3N^{'}}}\int \prod_{i=1}^{N^{'}}{ \mathrm{~d}^{3} \vec{q}_{i}}\exp \left[-\beta^{j}_{\mathrm{c}} \phi\left(\vec{q}_{1}, \cdots, \vec{q}_{N^{'}}\right)\right]\nonumber\\& \times \int \prod_{i=1}^{N^{'}}\left(\frac{\mathrm{d}^{3} \vec{p}_{i}}{h^{3}}\right)  \frac{1}{\mathcal{Z_{\textbf{Non-Int}}}(\beta^{j}_{\mathrm{c}}, N^{'}, V_{\mathrm{c}})} \exp \left[-\beta^{j}_{\mathrm{c}} \sum_{i=1}^{N^{'}} \frac{p_{i}^{2}}{2 m}\right]\int \prod_{i=1}^{N^{'}}\mathrm{~d}^{3} \vec{q}_{i}\nonumber\\&={\frac{1}{N^{'} !}\left[\frac{\sqrt{2 m k_{B} T^{j}_{\mathrm{c}}}}{h}\right]^{3N^{'}}}\int \prod_{i=1}^{N^{'}}\mathrm{~d}^{3} \vec{q}_{i}\nonumber\\& \times\int \prod_{i=1}^{N^{'}}{ \mathrm{~d}^{3} \vec{q}_{i}}\exp \left[-\beta^{j}_{\mathrm{c}} \phi\left(\vec{q}_{1}, \cdots, \vec{q}_{N^{'}}\right)\right] \nonumber\\&\times\int \prod_{i=1}^{N^{'}}\left(\frac{\mathrm{d}^{3} \vec{p}_{i}}{h^{3}}\right)  \frac{1}{\mathcal{Z_{\textbf{Non-Int}}}(\beta^{j}_{\mathrm{c}}, N^{'}, V_{\mathrm{c}})} \exp \left[-\beta^{j}_{\mathrm{c}} \sum_{i=1}^{N^{'}} \frac{p_{i}^{2}}{2 m}\right]\nonumber\\&
={\frac{1}{N^{'} !}\left[\frac{\sqrt{2 m k_{B} T^{j}_{\mathrm{c}}}}{h}\right]^{3N^{'}}}\int \prod_{i=1}^{N^{'}}\mathrm{~d}^{3} \vec{q}_{i}\nonumber\\& \times\underbrace{\int \prod_{i=1}^{N^{'}}\left(\frac{{\mathrm{~d}^{3} \vec{q}_{i}}\mathrm{d}^{3} \vec{p}_{i}}{h^{3}}\right) \frac{1}{\mathcal{Z_{\textbf{Non-Int}}}(\beta^{j}_{\mathrm{c}}, N^{'}, V_{\mathrm{c}})} \exp \left[-\beta^{j}_{\mathrm{c}} \sum_{i=1}^{N^{'}} \frac{p_{i}^{2}}{2 m}\right]\exp \left[-\beta^{j}_{\mathrm{c}} \phi\left(\vec{q}_{1}, \cdots, \vec{q}_{N^{'}}\right)\right]}_{\left\langle\exp \left[-\beta^{j}_{\mathrm{c}} \phi\left(\vec{q}_{1}, \cdots, \vec{q}_{N^{'}}\right)\right]\right\rangle_{\textbf{Non-Int}}}
\end{align}

\subsection{Attempt to write the weight distribution function without approximation (pseudo-Sophomore's dream)}
Let us assume that there is a possibility to derive an analytical form for $p\left(  \mu^{j}_{\mathrm{c}},\beta^{j},V_{\mathrm{c}}\right)$ without any approximation. By doing so, we can write 
\begin{align}
&p\left(  \mu^{j}_{\mathrm{c}},\beta^{j}_{\mathrm{c}},V_{\mathrm{c}}\right)={\mathcal{Q}^{-1}\left(g_1\left(\mu^{j}_{\mathrm{c}},\beta^{j}_{\mathrm{c}},V_{\mathrm{c}}\right),g_2\left(\mu^{j}_{\mathrm{c}},\beta^{j}_{\mathrm{c}},V_{\mathrm{c}}\right)+\ln{g_2\left(\mu^{j}_{\mathrm{c}},\beta^{j}_{\mathrm{c}},V_{\mathrm{c}}\right)}\right)}\nonumber\\&=\left[\sum_{N^{'}=0}^{\infty}
\mathrm{e}^{\beta^{j}_{\mathrm{c}} \mu^{j}_{\mathrm{c}} N^{'} } \mathcal{C}_{1}(\beta^{j}_{\mathrm{c}}, N^{'}, V_{\mathrm{c}})\nonumber\right.\\
&
\left.\exp\left[\left(\ln{g_2\left(\mu^{j}_{\mathrm{c}},\beta^{j}_{\mathrm{c}},V_{\mathrm{c}}\right)}+g_2\left(\mu^{j}_{\mathrm{c}},\beta^{j}_{\mathrm{c}},V_{\mathrm{c}}\right)\right)\, N^{'}-\left(\ln{g_2\left(\mu^{j}_{\mathrm{c}},\beta^{j}_{\mathrm{c}},V_{\mathrm{c}}\right)}+g_2\left(\mu^{j}_{\mathrm{c}},\beta^{j}_{\mathrm{c}},V_{\mathrm{c}}\right)\right)\right] \right]^{-1}\, ,
\end{align}
where 
\begin{align}\label{subnoapp}
\ln{g_2\left(\mu^{j}_{\mathrm{c}},\beta^{j}_{\mathrm{c}},V_{\mathrm{c}}\right)}+g_2\left(\mu^{j}_{\mathrm{c}},\beta^{j}_{\mathrm{c}},V_{\mathrm{c}}\right)=\ln\left[{\gamma\,k\,{e^{\beta^{j}_{\mathrm{c}}\mu^{j}_{\mathrm{c}}}}}\right]+\left[{\gamma\,k\,{e^{\beta^{j}_{\mathrm{c}}\mu^{j}_{\mathrm{c}}}}}\right]={\beta^{j}_{\mathrm{c}}\mu^{j}_{\mathrm{c}}}+\ln{\gamma\,k}+{\gamma\,k\,{e^{\beta^{j}_{\mathrm{c}}\mu^{j}_{\mathrm{c}}}}}\, .
\end{align}
With this we can write
\begin{align}
    &p\left(  \mu^{j}_{\mathrm{c}},\beta^{j}_{\mathrm{c}},V_{\mathrm{c}}\right)\nonumber\\&\propto \left[\sum_{N^{'}=0}^{\infty}
{\mathrm{e}^{\beta^{j}_{\mathrm{c}} \mu^{j}_{\mathrm{c}} N^{'} } \mathcal{C}_{1}(\beta^{j}_{\mathrm{c}}, N^{'}, V_{\mathrm{c}})\exp\left[({\beta^{j}_{\mathrm{c}}\mu^{j}_{\mathrm{c}}}+\ln{\gamma\,k}+{\gamma\,k\,{e^{\beta^{j}_{\mathrm{c}}\mu^{j}_{\mathrm{c}}}}})\, N^{'}-({\beta^{j}_{\mathrm{c}}\mu^{j}_{\mathrm{c}}}+\ln{\gamma\,k}+{\gamma\,k\,{e^{\beta^{j}_{\mathrm{c}}\mu^{j}_{\mathrm{c}}}}})\right]} \right]^{-1}\nonumber\\&
\propto \left[\sum_{N^{'}=0}^{\infty}
{\mathrm{e}^{\beta^{j}_{\mathrm{c}} \mu^{j}_{\mathrm{c}} N^{'} } \mathcal{C}_{1}(\beta^{j}_{\mathrm{c}}, N^{'}, V_{\mathrm{c}})\left[\,\mathrm{e}^{\beta^{j}_{\mathrm{c}}\mu^{j}_{\mathrm{c}}}\right]^{N^{'}}\left[\gamma\,k\right]^{N^{'}}\left[\mathrm{e}^{\gamma\,k\,{e^{\beta^{j}_{\mathrm{c}}\mu^{j}_{\mathrm{c}}}}}\right]^{N^{'}}\left[\,\mathrm{e}^{\beta^{j}_{\mathrm{c}}\mu^{j}_{\mathrm{c}}}\right]^{-1}\left[\gamma\,k\right]^{-1}\left[\mathrm{e}^{\gamma\,k\,{e^{\beta^{j}_{\mathrm{c}}\mu^{j}_{\mathrm{c}}}}}\right]^{-1}} \right]^{-1}
\nonumber\\&
\propto \left[\left[\mathrm{e}^{\beta^{j}_{\mathrm{c}}\mu^{j}_{\mathrm{c}}}\right]^{-1}\left[\gamma\,k\right]^{-1}\left[\mathrm{e}^{\gamma\,k\,{e^{\beta^{j}_{\mathrm{c}}\mu^{j}_{\mathrm{c}}}}}\right]^{-1}\sum_{N^{'}=0}^{\infty}\frac{1}{N^{'}!}
\left[\underbrace{{{V_{\mathrm{c}}}\left[\frac{\sqrt{2 m k_{B} T^{j}_{\mathrm{c}}}}{h}\right]^{3}}\,\mathrm{e}^{\beta^{j}_{\mathrm{c}}\mu^{j}_{\mathrm{c}}}\,\gamma\,k\,\mathrm{e}^{\gamma\,k\,{e^{\beta^{j}_{\mathrm{c}}\mu^{j}_{\mathrm{c}}}}}}_{\alpha}\right]^{N^{'}} \right]^{-1}
\end{align}
Interpreting the infinite sum over $N^{'}$as the power series of the exponential function,
\begin{align}
    \exp{\alpha}:=\sum_{N^{'}=0}^{\infty} \frac{\alpha^{N^{'}}}{N^{'} !} \, ,
\end{align}
 we arrive at
\begin{align}
   p\left(  \mu^{j}_{\mathrm{c}},\beta^{j}_{\mathrm{c}},V_{\mathrm{c}}\right)\propto\,\gamma\,k\,\mathrm{e}^{\beta^{j}_{\mathrm{c}}\mu^{j}_{\mathrm{c}}}\,\mathrm{e}^{\gamma\,k\,{e^{\beta^{j}_{\mathrm{c}}\mu^{j}_{\mathrm{c}}}}}\mathrm{e}^{{-\frac{\gamma\,k\,V_\mathrm{c}}{{\Lambda^{j}_{\mathrm{c}}}^{3}}}\mathrm{e}^{\beta^{j}_{\mathrm{c}} \mu^{j}_{\mathrm{c}}}\mathrm{e}^{\gamma\,k\,{e^{\beta^{j}_{\mathrm{c}}\mu^{j}_{\mathrm{c}}}}}}
\end{align}
we consider all artificial baths to have a fixed temperature equal to the thermal bath surrounding the composite system, i.e. $\beta^{j}_{\mathrm{c}} = \beta$, for all $j$ and only the artificial chemical potentials are allowed to fluctuate.
\begin{align}
 p\left(  \mu^{j}_{\mathrm{c}},\beta,V_{\mathrm{c}}\right)=\frac{\gamma\,k\,\mathrm{e}^{\beta\mu^{j}_{\mathrm{c}}}\,\mathrm{e}^{\gamma\,k\,{e^{\beta\mu^{j}_{\mathrm{c}}}}}\mathrm{e}^{{-\frac{\gamma\,k\,V_\mathrm{c}}{{\Lambda^{j}_{\mathrm{c}}}^{3}}}\mathrm{e}^{\beta\mu^{j}_{\mathrm{c}}}\mathrm{e}^{\gamma\,k\,{e^{\beta\mu^{j}_{\mathrm{c}}}}}}}{\sum_{j}\,\gamma\,k\,\mathrm{e}^{\beta\mu^{j}_{\mathrm{c}}}\,\mathrm{e}^{\gamma\,k\,{e^{\beta\mu^{j}_{\mathrm{c}}}}}\mathrm{e}^{{-\frac{\gamma\,k\,V_\mathrm{c}}{{\Lambda^{j}_{\mathrm{c}}}^{3}}}\mathrm{e}^{\beta \mu^{j}_{\mathrm{c}}}\mathrm{e}^{\gamma\,k\,{e^{\beta^{\mu^{j}_{\mathrm{c}}}}}}}}
\end{align}
a simplified notation with
\begin{align}
\quad\alpha_{1}=\,\gamma\,k,\quad
\alpha_{2}=\frac{\alpha_{1}\,V_\mathrm{c}}{{\Lambda^{j}_{\mathrm{c}}}^{3}},\quad \gamma>0
\end{align}
leads to
\begin{align}
        p\left(  \mu^{j}_{\mathrm{c}},\beta,V_{\mathrm{c}}\right)=\frac{\alpha_{1}\,\mathrm{e}^{\beta\mu^{j}_{\mathrm{c}}}\,\mathrm{e}^{\alpha_{1}\,{e^{\beta\mu^{j}_{\mathrm{c}}}}}\mathrm{e}^{-\alpha_{2}\,\mathrm{e}^{\beta \mu^{j}_{\mathrm{c}}}\mathrm{e}^{\alpha_{1}\,{e^{\beta\mu^{j}_{\mathrm{c}}}}}}}{\sum_{j}\,\alpha_{1}\,\mathrm{e}^{\beta\mu^{j}_{\mathrm{c}}}\,\mathrm{e}^{\alpha_{1}\,{e^{\beta\mu^{j}_{\mathrm{c}}}}}\mathrm{e}^{-\alpha_{2}\,\mathrm{e}^{\beta\mu^{j}_{\mathrm{c}}}\mathrm{e}^{\alpha_{1}\,{e^{\beta\mu^{j}_{\mathrm{c}}}}}}} \, .
\end{align}
The normalization factor of $p(  \mu^{j}_{\mathrm{c}},\beta,V_{\mathrm{c}})$ can be written as
\begin{align}
{\sum_{j}\,\alpha_{1}\,\mathrm{e}^{\beta\mu^{j}_{\mathrm{c}}}\,\mathrm{e}^{\alpha_{1}\,{e^{\beta\mu^{j}_{\mathrm{c}}}}}\mathrm{e}^{-\alpha_{2}\,\mathrm{e}^{\beta \mu^{j}_{\mathrm{c}}}\mathrm{e}^{\alpha_{1}\,{e^{\beta\mu^{j}_{\mathrm{c}}}}}}}   =\beta\int_{-\infty}^{\infty} \text{d}\mu_{\mathrm{c}}\,\alpha_{1}\,\mathrm{e}^{\beta\mu_{\mathrm{c}}}\,\mathrm{e}^{\alpha_{1}\,{e^{\beta\mu_{\mathrm{c}}}}}\mathrm{e}^{-\alpha_{2}\,\mathrm{e}^{\beta \mu_{\mathrm{c}}}\mathrm{e}^{\alpha_{1}\,{e^{\beta\mu_{\mathrm{c}}}}}}\, .
\end{align}
We now impose a change of variables $\mathrm{e}^{\alpha_{1}\,{e^{\beta\mu_{\mathrm{c}}}}}=u $, such that $ \alpha_{1}\,{e^{\beta\mu_{\mathrm{c}}}} \text{d}\mu_{\mathrm{c}} ={\text{d}\,u}/{\beta\,u} $ and ${e^{\beta\mu_{\mathrm{c}}}}= {\ln{u}}/{\alpha_1}$. We also note that for a hard sphere system of particles we can easily prove that $k < 0$ and on the other hand we know $\gamma > 0$ then $\alpha_1 < 0$.
\begin{align}
   &\beta\int_{-\infty}^{\infty} \text{d}\mu_{\mathrm{c}}\,\alpha_{1}\,\mathrm{e}^{\beta\mu_{\mathrm{c}}}\,\mathrm{e}^{\alpha_{1}\,{e^{\beta\mu_{\mathrm{c}}}}}\mathrm{e}^{-\alpha_{2}\,\mathrm{e}^{\beta \mu_{\mathrm{c}}}\mathrm{e}^{\alpha_{1}\,{e^{\beta\mu_{\mathrm{c}}}}}}= -\beta\int_{0}^{1} \, \frac{\text{d}u}{\beta}\,u^{-1}\,u^{1}\,\mathrm{e}^{-\frac{\alpha_{2}}{\alpha_1}\,u\,\ln{u}}\nonumber\\&= -
   \frac{\beta}{\beta}\int_{0}^{1} \, {\text{d}u}\,\mathrm{e}^{-\frac{\alpha_{2}}{\alpha_1}\,u\,\ln{u}}= -
   \int_{0}^{1} \, {\text{d}u}\,\sum_{n=0}^{\infty} ({-\frac{\alpha_2}{\alpha_1}})^{n}\frac{u^{n}(\ln{u})^{n}}{n !} \nonumber\\&=-\sum_{n=0}^{\infty}\,({-\frac{\alpha_2}{\alpha_1}})^{n}\,\int_{0}^{1} {\text{d}u}\frac{u^{n}(\ln{u})^{n}}{n !}
\end{align}
The above integral was historically computed by Johann Bernoulli \cite{dunham2005touring}, 
\begin{align}
    \int_{0}^{1}{\text{d}u} \frac{u^{n}(\ln{u})^{n}}{n !} = (-1)^{n}(n+1)^{-(n+1)}\, .
\end{align}
With this we arrive at
\begin{align}
    {\beta}\int_{-\infty}^{\infty} \text{d}\mu_{\mathrm{c}}\,\alpha_{1}\,\mathrm{e}^{\beta\mu_{\mathrm{c}}}\,\mathrm{e}^{\alpha_{1}\,{e^{\beta\mu_{\mathrm{c}}}}}\mathrm{e}^{-\alpha_{2}\,\mathrm{e}^{\beta \mu_{\mathrm{c}}}\mathrm{e}^{\alpha_{1}\,{e^{\beta\mu_{\mathrm{c}}}}}}= -\sum_{n=0}^{\infty} \left(\frac{V_\mathrm{c}}{{\Lambda_{\mathrm{c}}}^{3}}\right)^{n}(n+1)^{-(n+1)} \, .
\end{align}
The infinite sum can only be computed when the inter-particle distance approaches the radius of the solvation shell, i.e. ${V_\mathrm{c}}/{{\Lambda_{\mathrm{c}}}^{3}} = 1$. Only under this condition we can determine the normalization as
\begin{align}
{\sum_{j}\,\alpha_{1}\,\mathrm{e}^{\beta\mu^{j}_{\mathrm{c}}}\,\mathrm{e}^{\alpha_{1}\,{e^{\beta\mu^{j}_{\mathrm{c}}}}}\mathrm{e}^{-\alpha_{2}\,\mathrm{e}^{\beta \mu^{j}_{\mathrm{c}}}\mathrm{e}^{\alpha_{1}\,{e^{\beta\mu^{j}_{\mathrm{c}}}}}}} =-\sum_{n=0}^{\infty} (n+1)^{-(n+1)}\approx\,-1.29 \, ,
\end{align}
and write the weighted distribution function as
\begin{align}
p\left(  \mu^{j}_{\mathrm{c}},\beta,V_{\mathrm{c}}\right)=\frac{\alpha_{1}\,\mathrm{e}^{\beta\mu^{j}_{\mathrm{c}}}\,\mathrm{e}^{\alpha_{1}\,{e^{\beta\mu^{j}_{\mathrm{c}}}}}\mathrm{e}^{-\alpha_{1}\,\mathrm{e}^{\beta \mu^{j}_{\mathrm{c}}}\mathrm{e}^{\alpha_{1}\,{e^{\beta\mu^{j}_{\mathrm{c}}}}}}}{-1.29}\quad \text{s.t.}\quad\alpha_{1}=\,\gamma\,k<0,\quad \gamma>0 \, . 
\end{align}
Since this condition for which we can solve the expression is unphysical, we decided to approximate $g_2 + \ln g_2$ by introducing the two parameters $\lambda_1$ and $\lambda_2$ as shown in Eq. \ref{taylor} and arrive at a model for which we can formulate an analytic form without imposing unphysical conditions.
\subsection{Moments in terms of the cumulants}\cite{kardar2007statistical}
 The expectation value of $\mathrm{e}^{-\mathrm{i} k x}$ can be written as,
 \begin{align}\label{eq2}
 \left\langle\mathrm{e}^{-\mathrm{i} k x}\right\rangle=\int \mathrm{d} x f(x) \mathrm{e}^{-\mathrm{i} k x} =\mathcal{F}_{x}[f(x)](k)=F(k) \, ,
 \end{align}
 in which $f(x)$ is the density probability function. We can interpret $F(k)$ as the Fourier transform of the probability density function $f(x)$, called the characteristic function.
 If we expand $\mathrm{e}^{-\mathrm{i} k x}$
 based its power series,
  \begin{align}\label{eq3}
 \left\langle\mathrm{e}^{-\mathrm{i} k x}\right\rangle=\left\langle\sum_{m=0}^{\infty} \frac{(-\mathrm{i} k)^{m}}{m !} x^{m}\right\rangle=\sum_{m=0}^{\infty} \frac{(-\mathrm{i} k)^{m}}{m !}\left\langle x^{m}\right\rangle\, .\end{align}
 The logarithm of the characteristic function, $F(k)$,  is the cumulant generating function. Expanding this expression gives the cumulants of the distribution defined by
 \begin{align}
     \ln F(k)=\sum_{m=1}^{\infty} \frac{(-\mathrm{i} k)^{m}}{m !}\left\langle x^{m}\right\rangle_{c}
 \end{align}
 and hence
 \begin{align}\label{eq4}
     F(k)=\exp\left[\sum_{m=1}^{\infty} \frac{(-\mathrm{i} k)^{m}}{m !}\left\langle x^{m}\right\rangle_{c}\right] \, .
 \end{align} 
Combining the Eqns.~(\ref{eq2}), (\ref{eq3}) and (\ref{eq4}), we obtain
 \begin{align}\label{eq6}
     \exp\left[\sum_{m=1}^{\infty} \frac{(-\mathrm{i} k)^{m}}{m !}\left\langle x^{m}\right\rangle_{c}\right]=\sum_{m=0}^{\infty} \frac{(-\mathrm{i} k)^{m}}{m !}\left\langle x^{m}\right\rangle \, .
 \end{align}

\begin{acknowledgement}

The authors thank the Deutsche Forschungsgemeinschaft (DFG, German Research Foundation) under Germany´s Excellence Strategy – EXC 2033 – 390677874 – RESOLV for funding. C. J. S. further acknowledges funding by the Ministry of Innovation, Science and Research of North Rhine-Westphalia (“NRW Rückkehrerprogramm”).
M. R. is grateful to Dipesh Somvanshi and Siyavash Moradi for help with the implementation of the algorithms.
The authors also thank Prof. Martin Head-Gordon for introducing them to the work of Prof. Dilip Asthagiri and hence initiating this study.

\end{acknowledgement}

\bibliography{article_final}

\providecommand{\latin}[1]{#1}
\makeatletter
\providecommand{\doi}
  {\begingroup\let\do\@makeother\dospecials
  \catcode`\{=1 \catcode`\}=2 \doi@aux}
\providecommand{\doi@aux}[1]{\endgroup\texttt{#1}}
\makeatother
\providecommand*\mcitethebibliography{\thebibliography}
\csname @ifundefined\endcsname{endmcitethebibliography}
  {\let\endmcitethebibliography\endthebibliography}{}
\begin{mcitethebibliography}{35}
\providecommand*\natexlab[1]{#1}
\providecommand*\mciteSetBstSublistMode[1]{}
\providecommand*\mciteSetBstMaxWidthForm[2]{}
\providecommand*\mciteBstWouldAddEndPuncttrue
  {\def\EndOfBibitem{\unskip.}}
\providecommand*\mciteBstWouldAddEndPunctfalse
  {\let\EndOfBibitem\relax}
\providecommand*\mciteSetBstMidEndSepPunct[3]{}
\providecommand*\mciteSetBstSublistLabelBeginEnd[3]{}
\providecommand*\EndOfBibitem{}
\mciteSetBstSublistMode{f}
\mciteSetBstMaxWidthForm{subitem}{(\alph{mcitesubitemcount})}
\mciteSetBstSublistLabelBeginEnd
  {\mcitemaxwidthsubitemform\space}
  {\relax}
  {\relax}

\bibitem[Sethna(2021)]{sethna2021statistical}
Sethna,~J. \emph{Statistical mechanics: entropy, order parameters, and
  complexity}; Oxford University Press, USA, 2021; Vol.~14\relax
\mciteBstWouldAddEndPuncttrue
\mciteSetBstMidEndSepPunct{\mcitedefaultmidpunct}
{\mcitedefaultendpunct}{\mcitedefaultseppunct}\relax
\EndOfBibitem
\bibitem[Mandelbrot(1989)]{mandelbrot1989temperature}
Mandelbrot,~B.~B. Temperature fluctuation: a well-defined and unavoidable
  notion. \emph{Phys. Today} \textbf{1989}, \emph{42}, 71\relax
\mciteBstWouldAddEndPuncttrue
\mciteSetBstMidEndSepPunct{\mcitedefaultmidpunct}
{\mcitedefaultendpunct}{\mcitedefaultseppunct}\relax
\EndOfBibitem
\bibitem[Dunkel and Hilbert(2014)Dunkel, and Hilbert]{dunkel2014consistent}
Dunkel,~J.; Hilbert,~S. Consistent thermostatistics forbids negative absolute
  temperatures. \emph{Nat. Phys.} \textbf{2014}, \emph{10}, 67--72\relax
\mciteBstWouldAddEndPuncttrue
\mciteSetBstMidEndSepPunct{\mcitedefaultmidpunct}
{\mcitedefaultendpunct}{\mcitedefaultseppunct}\relax
\EndOfBibitem
\bibitem[Mandelbrot(1962)]{mandelbrot1962role}
Mandelbrot,~B. The role of sufficiency and of estimation in thermodynamics.
  \emph{Ann. Math. Stat.} \textbf{1962}, 1021--1038\relax
\mciteBstWouldAddEndPuncttrue
\mciteSetBstMidEndSepPunct{\mcitedefaultmidpunct}
{\mcitedefaultendpunct}{\mcitedefaultseppunct}\relax
\EndOfBibitem
\bibitem[Dixit(2015)]{dixit2015detecting}
Dixit,~P.~D. Detecting temperature fluctuations at equilibrium. \emph{PCCP}
  \textbf{2015}, \emph{17}, 13000--13005\relax
\mciteBstWouldAddEndPuncttrue
\mciteSetBstMidEndSepPunct{\mcitedefaultmidpunct}
{\mcitedefaultendpunct}{\mcitedefaultseppunct}\relax
\EndOfBibitem
\bibitem[Pratt and Rempe(1999)Pratt, and Rempe]{pratt1999quasi}
Pratt,~L.~R.; Rempe,~S.~B. Quasi-chemical theory and implicit solvent models
  for simulations. AIP Conf. Proc. 1999; pp 172--201\relax
\mciteBstWouldAddEndPuncttrue
\mciteSetBstMidEndSepPunct{\mcitedefaultmidpunct}
{\mcitedefaultendpunct}{\mcitedefaultseppunct}\relax
\EndOfBibitem
\bibitem[Juffer and Berendsen(1993)Juffer, and Berendsen]{juffer1993dynamic}
Juffer,~A.; Berendsen,~H. Dynamic surface boundary conditions: A simple
  boundary model for molecular dynamics simulations. \emph{Mol. Phys.}
  \textbf{1993}, \emph{79}, 623--644\relax
\mciteBstWouldAddEndPuncttrue
\mciteSetBstMidEndSepPunct{\mcitedefaultmidpunct}
{\mcitedefaultendpunct}{\mcitedefaultseppunct}\relax
\EndOfBibitem
\bibitem[Remsing \latin{et~al.}(2016)Remsing, Liu, and Weeks]{remsing2016}
Remsing,~R.~C.; Liu,~S.; Weeks,~J.~D. Long-ranged contributions to solvation
  free energies from theory and short-ranged models. \emph{Proc. Natl. Acad.
  Sci. U.S.A.} \textbf{2016}, \emph{113}, 2819--2826\relax
\mciteBstWouldAddEndPuncttrue
\mciteSetBstMidEndSepPunct{\mcitedefaultmidpunct}
{\mcitedefaultendpunct}{\mcitedefaultseppunct}\relax
\EndOfBibitem
\bibitem[Hegemann \latin{et~al.}(2017)Hegemann, Hocquard, and
  Heuberger]{hegemann2017}
Hegemann,~D.; Hocquard,~N.; Heuberger,~M. Nanoconfined water can orient and
  cause long-range dipolar interactions with biomolecules. \emph{Sci. Rep.}
  \textbf{2017}, \emph{7}, 2045--2322\relax
\mciteBstWouldAddEndPuncttrue
\mciteSetBstMidEndSepPunct{\mcitedefaultmidpunct}
{\mcitedefaultendpunct}{\mcitedefaultseppunct}\relax
\EndOfBibitem
\bibitem[Belch and Berkowitz(1985)Belch, and Berkowitz]{belch1985molecular}
Belch,~A.~C.; Berkowitz,~M. Molecular dynamics simulations of TIPS2 water
  restricted by a spherical hydrophobic boundary. \emph{Chem. Phys. Lett.}
  \textbf{1985}, \emph{113}, 278--282\relax
\mciteBstWouldAddEndPuncttrue
\mciteSetBstMidEndSepPunct{\mcitedefaultmidpunct}
{\mcitedefaultendpunct}{\mcitedefaultseppunct}\relax
\EndOfBibitem
\bibitem[Klamt \latin{et~al.}(1998)Klamt, Jonas, Bürger, and
  Lohrenz]{klamt1989}
Klamt,~A.; Jonas,~V.; Bürger,~T.; Lohrenz,~J. C.~W. Refinement and
  Parametrization of COSMO-RS. \emph{J. Phys. Chem. A} \textbf{1998},
  \emph{102}, 5074--5085\relax
\mciteBstWouldAddEndPuncttrue
\mciteSetBstMidEndSepPunct{\mcitedefaultmidpunct}
{\mcitedefaultendpunct}{\mcitedefaultseppunct}\relax
\EndOfBibitem
\bibitem[Sato \latin{et~al.}(2000)Sato, Kovalenko, and Hirata]{sato2000}
Sato,~H.; Kovalenko,~A.; Hirata,~F. Self-consistent field, ab initio molecular
  orbital and three-dimensional reference interaction site model study for
  solvation effect on carbon monoxide in aqueous solution. \emph{J. Chem.
  Phys.} \textbf{2000}, \emph{112}, 9463--9468\relax
\mciteBstWouldAddEndPuncttrue
\mciteSetBstMidEndSepPunct{\mcitedefaultmidpunct}
{\mcitedefaultendpunct}{\mcitedefaultseppunct}\relax
\EndOfBibitem
\bibitem[Mennucci(2010)]{mennucci2010}
Mennucci,~B. Continuum Solvation Models: What Else Can We Learn from Them?
  \emph{J, Phys. Chem. Lett.} \textbf{2010}, \emph{1}, 1666--1674\relax
\mciteBstWouldAddEndPuncttrue
\mciteSetBstMidEndSepPunct{\mcitedefaultmidpunct}
{\mcitedefaultendpunct}{\mcitedefaultseppunct}\relax
\EndOfBibitem
\bibitem[Stein \latin{et~al.}(2019)Stein, Herbert, and Head-Gordon]{stein2019}
Stein,~C.~J.; Herbert,~J.~M.; Head-Gordon,~M. The Poisson--Boltzmann model for
  implicit solvation of electrolyte solutions: Quantum chemical implementation
  and assessment via Sechenov coefficients. \emph{J. Chem. Phys.}
  \textbf{2019}, \emph{151}, 224111\relax
\mciteBstWouldAddEndPuncttrue
\mciteSetBstMidEndSepPunct{\mcitedefaultmidpunct}
{\mcitedefaultendpunct}{\mcitedefaultseppunct}\relax
\EndOfBibitem
\bibitem[Herbert(2021)]{herbert2021}
Herbert,~J.~M. Dielectric continuum methods for quantum chemistry. \emph{WIREs
  Comput. Mol. Sci.} \textbf{2021}, \emph{11}, e1519\relax
\mciteBstWouldAddEndPuncttrue
\mciteSetBstMidEndSepPunct{\mcitedefaultmidpunct}
{\mcitedefaultendpunct}{\mcitedefaultseppunct}\relax
\EndOfBibitem
\bibitem[Ringe \latin{et~al.}(2022)Ringe, Hörmann, Oberhofer, and
  Reuter]{ringe2022}
Ringe,~S.; Hörmann,~N.~G.; Oberhofer,~H.; Reuter,~K. Implicit Solvation
  Methods for Catalysis at Electrified Interfaces. \emph{Chem. Rev.}
  \textbf{2022}, \emph{122}, 10777--10820\relax
\mciteBstWouldAddEndPuncttrue
\mciteSetBstMidEndSepPunct{\mcitedefaultmidpunct}
{\mcitedefaultendpunct}{\mcitedefaultseppunct}\relax
\EndOfBibitem
\bibitem[Thomas \latin{et~al.}(2013)Thomas, Chun, Chen, Wei, and
  Baker]{thomas2013parameterization}
Thomas,~D.~G.; Chun,~J.; Chen,~Z.; Wei,~G.; Baker,~N.~A. Parameterization of a
  geometric flow implicit solvation model. \emph{J. Comput. Chem.}
  \textbf{2013}, \emph{34}, 687--695\relax
\mciteBstWouldAddEndPuncttrue
\mciteSetBstMidEndSepPunct{\mcitedefaultmidpunct}
{\mcitedefaultendpunct}{\mcitedefaultseppunct}\relax
\EndOfBibitem
\bibitem[Lange \latin{et~al.}(2020)Lange, Herbert, Albrecht, and
  You]{lange2020}
Lange,~A.~W.; Herbert,~J.~M.; Albrecht,~B.~J.; You,~Z.-Q. Intrinsically smooth
  discretisation of Connolly's solvent-excluded molecular surface. \emph{Mol.
  Phys.} \textbf{2020}, \emph{118}, e1644384\relax
\mciteBstWouldAddEndPuncttrue
\mciteSetBstMidEndSepPunct{\mcitedefaultmidpunct}
{\mcitedefaultendpunct}{\mcitedefaultseppunct}\relax
\EndOfBibitem
\bibitem[Rana and Nguyen(2022)Rana, and Nguyen]{rana2022}
Rana,~M.~M.; Nguyen,~D.~D. EISA-Score: Element Interactive Surface Area Score
  for Protein–Ligand Binding Affinity Prediction. \emph{J. Chem. Inf.
  Model.J} \textbf{2022}, \emph{62}, 4329--4341\relax
\mciteBstWouldAddEndPuncttrue
\mciteSetBstMidEndSepPunct{\mcitedefaultmidpunct}
{\mcitedefaultendpunct}{\mcitedefaultseppunct}\relax
\EndOfBibitem
\bibitem[Li and Hartke(2013)Li, and Hartke]{li2013}
Li,~Y.; Hartke,~B. Assessing Solvation Effects on Chemical Reactions with
  Globally Optimized Solvent Clusters. \emph{Chem. Phys. Chem.} \textbf{2013},
  \emph{14}, 2678--2686\relax
\mciteBstWouldAddEndPuncttrue
\mciteSetBstMidEndSepPunct{\mcitedefaultmidpunct}
{\mcitedefaultendpunct}{\mcitedefaultseppunct}\relax
\EndOfBibitem
\bibitem[Kildgaard \latin{et~al.}(2018)Kildgaard, Mikkelsen, Bilde, and
  Elm]{kildgaard2018}
Kildgaard,~J.~V.; Mikkelsen,~K.~V.; Bilde,~M.; Elm,~J. Hydration of Atmospheric
  Molecular Clusters: A New Method for Systematic Configurational Sampling.
  \emph{J. Phys. Chem. A} \textbf{2018}, \emph{122}, 5026--5036\relax
\mciteBstWouldAddEndPuncttrue
\mciteSetBstMidEndSepPunct{\mcitedefaultmidpunct}
{\mcitedefaultendpunct}{\mcitedefaultseppunct}\relax
\EndOfBibitem
\bibitem[Simm \latin{et~al.}(2020)Simm, Türtscher, and Reiher]{simm2020}
Simm,~G.~N.; Türtscher,~P.~L.; Reiher,~M. Systematic microsolvation approach
  with a cluster-continuum scheme and conformational sampling. \emph{J. Comput.
  Chem.} \textbf{2020}, \emph{41}, 1144--1155\relax
\mciteBstWouldAddEndPuncttrue
\mciteSetBstMidEndSepPunct{\mcitedefaultmidpunct}
{\mcitedefaultendpunct}{\mcitedefaultseppunct}\relax
\EndOfBibitem
\bibitem[Steiner \latin{et~al.}(2021)Steiner, Holzknecht, Schauperl, and
  Podewitz]{steiner2021}
Steiner,~M.; Holzknecht,~T.; Schauperl,~M.; Podewitz,~M. Quantum Chemical
  Microsolvation by Automated Water Placement. \emph{Molecules} \textbf{2021},
  \emph{26}\relax
\mciteBstWouldAddEndPuncttrue
\mciteSetBstMidEndSepPunct{\mcitedefaultmidpunct}
{\mcitedefaultendpunct}{\mcitedefaultseppunct}\relax
\EndOfBibitem
\bibitem[Dixit \latin{et~al.}(2017)Dixit, Bansal, Chapman, and
  Asthagiri]{dixit2017mini}
Dixit,~P.~D.; Bansal,~A.; Chapman,~W.~G.; Asthagiri,~D. Mini-grand canonical
  ensemble: chemical potential in the solvation shell. \emph{J. Chem. Phys.}
  \textbf{2017}, \emph{147}, 164901\relax
\mciteBstWouldAddEndPuncttrue
\mciteSetBstMidEndSepPunct{\mcitedefaultmidpunct}
{\mcitedefaultendpunct}{\mcitedefaultseppunct}\relax
\EndOfBibitem
\bibitem[Dixit(2013)]{dixit2013maximum}
Dixit,~P.~D. A maximum entropy thermodynamics of small systems. \emph{J. Chem.
  Phys.} \textbf{2013}, \emph{138}, 184111\relax
\mciteBstWouldAddEndPuncttrue
\mciteSetBstMidEndSepPunct{\mcitedefaultmidpunct}
{\mcitedefaultendpunct}{\mcitedefaultseppunct}\relax
\EndOfBibitem
\bibitem[Bansal \latin{et~al.}(2017)Bansal, Valiya~Parambathu, Asthagiri, Cox,
  and Chapman]{bansal2017thermodynamics}
Bansal,~A.; Valiya~Parambathu,~A.; Asthagiri,~D.; Cox,~K.~R.; Chapman,~W.~G.
  Thermodynamics of mixtures of patchy and spherical colloids of different
  sizes: A multi-body association theory with complete reference fluid
  information. \emph{J. Chem. Phys.} \textbf{2017}, \emph{146}, 164904\relax
\mciteBstWouldAddEndPuncttrue
\mciteSetBstMidEndSepPunct{\mcitedefaultmidpunct}
{\mcitedefaultendpunct}{\mcitedefaultseppunct}\relax
\EndOfBibitem
\bibitem[Bansal \latin{et~al.}(2016)Bansal, Asthagiri, Cox, and
  Chapman]{bansal2016structure}
Bansal,~A.; Asthagiri,~D.; Cox,~K.~R.; Chapman,~W.~G. Structure and
  thermodynamics of a mixture of patchy and spherical colloids: A multi-body
  association theory with complete reference fluid information. \emph{J. Chem.
  Phys.} \textbf{2016}, \emph{145}, 074904\relax
\mciteBstWouldAddEndPuncttrue
\mciteSetBstMidEndSepPunct{\mcitedefaultmidpunct}
{\mcitedefaultendpunct}{\mcitedefaultseppunct}\relax
\EndOfBibitem
\bibitem[Kardar(2007)]{kardar2007statistical}
Kardar,~M. \emph{Statistical physics of particles}; Cambridge University Press,
  2007\relax
\mciteBstWouldAddEndPuncttrue
\mciteSetBstMidEndSepPunct{\mcitedefaultmidpunct}
{\mcitedefaultendpunct}{\mcitedefaultseppunct}\relax
\EndOfBibitem
\bibitem[Tetrode(1912)]{tetrode1912chemische}
Tetrode,~H. Die chemische Konstante der Gase und das elementare
  Wirkungsquantum. \emph{Annalen der Physik} \textbf{1912}, \emph{343},
  434--442\relax
\mciteBstWouldAddEndPuncttrue
\mciteSetBstMidEndSepPunct{\mcitedefaultmidpunct}
{\mcitedefaultendpunct}{\mcitedefaultseppunct}\relax
\EndOfBibitem
\bibitem[Sackur(1911)]{sackur1911anwendung}
Sackur,~O. Die Anwendung der kinetischen Theorie der Gase auf chemische
  Probleme. \emph{Annalen der Physik} \textbf{1911}, \emph{341}, 958--980\relax
\mciteBstWouldAddEndPuncttrue
\mciteSetBstMidEndSepPunct{\mcitedefaultmidpunct}
{\mcitedefaultendpunct}{\mcitedefaultseppunct}\relax
\EndOfBibitem
\bibitem[Beck \latin{et~al.}(2006)Beck, Paulaitis, and
  Pratt]{beck2006potential}
Beck,~T.~L.; Paulaitis,~M.~E.; Pratt,~L.~R. \emph{The potential distribution
  theorem and models of molecular solutions}; Cambridge University Press,
  2006\relax
\mciteBstWouldAddEndPuncttrue
\mciteSetBstMidEndSepPunct{\mcitedefaultmidpunct}
{\mcitedefaultendpunct}{\mcitedefaultseppunct}\relax
\EndOfBibitem
\bibitem[Widom(1963)]{widom1963some}
Widom,~B. Some topics in the theory of fluids. \emph{J. Chem. Phys.}
  \textbf{1963}, \emph{39}, 2808--2812\relax
\mciteBstWouldAddEndPuncttrue
\mciteSetBstMidEndSepPunct{\mcitedefaultmidpunct}
{\mcitedefaultendpunct}{\mcitedefaultseppunct}\relax
\EndOfBibitem
\bibitem[Rowlinson and Widom(1982)Rowlinson, and Widom]{rowlinson1982molecular}
Rowlinson,~J.; Widom,~B. Molecular Theory of Capillarity (Oxford University
  Press). 1982\relax
\mciteBstWouldAddEndPuncttrue
\mciteSetBstMidEndSepPunct{\mcitedefaultmidpunct}
{\mcitedefaultendpunct}{\mcitedefaultseppunct}\relax
\EndOfBibitem
\bibitem[Dunham(2005)]{dunham2005touring}
Dunham,~W. Touring the calculus gallery. \emph{Am. Math. Mon.} \textbf{2005},
  \emph{112}, 1--19\relax
\mciteBstWouldAddEndPuncttrue
\mciteSetBstMidEndSepPunct{\mcitedefaultmidpunct}
{\mcitedefaultendpunct}{\mcitedefaultseppunct}\relax
\EndOfBibitem
\end{mcitethebibliography}

\end{document}